\pdfoutput=1
\documentclass[aps,prd,amsmath,floats,floatfix, twocolumn,
superscriptaddress,nofootinbib,showpacs]{revtex4-1}

\usepackage[T1]{fontenc}
\usepackage[utf8]{inputenc}
\usepackage{lmodern}
\usepackage{subfigure}
\usepackage{verbatim}

\usepackage[dvipsnames, usenames]{xcolor}
\definecolor{linkcolor}{rgb}{0.0,0.3,0.5}
\usepackage[hypertexnames=false, unicode, colorlinks=true, linkcolor=linkcolor,
citecolor=linkcolor, filecolor=linkcolor,urlcolor=linkcolor,
pdfusetitle]{hyperref}

\usepackage[all]{hypcap}
\usepackage{graphicx}
\usepackage{xspace}
\usepackage{amssymb}
\usepackage[normalem]{ulem} %for \sout
\usepackage{multirow}
\usepackage{graphicx}
\usepackage{gensymb}
\usepackage{microtype}

\usepackage[english]{babel}
\usepackage{blindtext}

\graphicspath{%
	{figs/}%
}

\DeclareMathAlphabet{\mathpzc}{OT1}{pzc}{m}{it}

\newcommand{\chieff}{\chi_{\mathrm{eff}}}

\begin{document}
	
	\title{Improved analysis of GW190412 with a\\
		precessing numerical relativity surrogate waveform model}
	
	\newcommand{\LIGOlabMIT}{\affiliation{LIGO Laboratory, Massachusetts Institute of Technology, 185 Albany St, Cambridge, MA 02139, USA}}
       \newcommand{\MKI}{\affiliation{Department of Physics and Kavli Institute for Astrophysics and Space Research, Massachusetts Institute of Technology, 77 Massachusetts Ave, Cambridge, MA 02139, USA}}
	\newcommand{\monash}{\affiliation{School of Physics and Astronomy, Monash University, Vic 3800, Australia}}
	\newcommand{\ozgrav}{\affiliation{OzGrav: The ARC Centre of Excellence for Gravitational Wave Discovery, Clayton VIC 3800, Australia}}
	\newcommand{\UMassDMath}{\affiliation{Department of Mathematics and
			Center for Scientific Computing and Visualization Research,
			University of Massachusetts, Dartmouth, MA 02747, USA}} 
	\newcommand{\UMassDPhy}{\affiliation{Department of Physics and
			Center for Scientific Computing and Visualization Research,
			University of Massachusetts, Dartmouth, MA 02747, USA}} 
	
	\author{Tousif Islam}
	\email{tislam@umassd.edu}
	\UMassDPhy
	\author{Scott E. Field}
	\email{sfield@umassd.edu}
	\UMassDMath
	\author{Carl-Johan Haster}
	\email{haster@mit.edu}
	\LIGOlabMIT
	\MKI
	\author{Rory Smith}
	\email{rory.smith@monash.edu}
	\monash
	\ozgrav

	% Because hyperref only gets the *last* author, we need to be explicit.
	\hypersetup{pdfauthor={Islam et al.}}
	
	\date{\today}
	
	%==========================================================================
	\begin{abstract}
		The recent observation of GW190412, the first high-mass ratio binary black-hole (BBH) merger, by the LIGO-Virgo Collaboration (LVC) provides a unique opportunity to probe the impact of subdominant harmonics and precession effects encoded in a gravitational wave signal. 
		We present refined estimates of source parameters for GW190412 using \texttt{NRSur7dq4}, a recently developed numerical relativity waveform surrogate model that includes all $\ell \leq 4$ spin-weighted spherical harmonic modes as well as the  full physical effects of precession.
		We compare our results with two different variants of phenomenological precessing BBH waveform models, \texttt{IMRPhenomPv3HM} and \texttt{IMRPhenomXPHM}, as well as to the LVC results. 
		Our results are broadly in agreement with \texttt{IMRPhenomXPHM} results and the reported LVC analysis compiled with the \texttt{SEOBNRv4PHM} waveform model, but in tension with \texttt{IMRPhenomPv3HM}. 
		Using the \texttt{NRSur7dq4} model, we provide a tighter constraint on the mass-ratio ($0.26^{+0.08}_{-0.06}$) as compared to the LVC estimate of $0.28^{+0.13}_{-0.07}$ (both reported as median values withs 90\% credible intervals).
		We also constrain the binary to be more face-on, and find a broader posterior for the spin precession parameter.
		We further find that even though $\ell=4$ harmonic modes have negligible signal-to-noise ratio, omission of these modes will influence the estimated posterior distribution of several source parameters including chirp mass, effective inspiral spin, luminosity distance, and inclination. 
		We also find that commonly used model approximations, such as neglecting the asymmetric modes (which are generically excited during precession), have negligible impact on parameter recovery for moderate SNR-events similar to GW190412.
	\end{abstract}
	
	\maketitle
	%==========================================================================
	%==========================================================================
	%==========================================================================
	\section{Introduction} 
	\label{Sec:Introduction}
	
	The observation of gravitational waves (GWs) in the LIGO and Virgo detectors~\cite{Harry:2010zz, TheVirgo:2014hva} from the merger of two black holes (BHs) of unequal mass was recently reported by the LIGO-Virgo Collaboration (LVC)~\cite{ligo2020gw190412}.
	This observation, GW190412, is unique in the sense that it is not only the first asymmetric mass-ratio event, but it is also the first event with a strong constraint on the spin magnitude associated with the more massive BH.
	The original LVC analysis~\cite{ligo2020gw190412} also finds moderate support for spin-induced orbital precession. 
	Such high spin asymmetric mass-ratio systems excite several higher order harmonics of the waveform.
	GW190412 therefore provides an excellent opportunity to probe beyond the dominant multipole of the GW signal.
	
	Data from LIGO and Virgo containing GW190412~\cite{GWOSC_GW190412} was subsequently matched against several state-of-art precessing GW signal model approximants in order to extract the source parameters. 
	In the original analysis from the LIGO-Virgo Collaboration (LVC), this includes approximants from both the effective-one-body (EOB) waveform family (\texttt{SEOBNRv4\_ROM} \cite{bohe2017improved}, \texttt{SEOBNRv4HM\_ROM} \cite{cotesta2018enriching}, \texttt{SEOBNRv4P} \cite{cotesta2020frequency, pan2014inspiral,babak2017validating} and \texttt{SEOBNRv4PHM} \cite{cotesta2020frequency, pan2014inspiral,babak2017validating}) and the phenomenological (Phenom) waveform family (\texttt{IMRPhenomD} \cite{husa2016frequency,khan2016frequency}, \texttt{IMRPhenomHM} \cite{london2018first}, \texttt{IMRPhenomPv2} \cite{khan2019phenomenological,hannam2014simple}, \texttt{IMRPhenomPv3HM} \cite{khan2020including}), and a numerical relativity (NR) surrogate waveform for aligned spin GW signals \texttt{NRHybSur3dq8} \cite{varma2019surrogate2}. 
	The inferred parameter values for GW190412 quoted by the LVC used \texttt{SEOBNRv4PHM} and \texttt{IMRPhenomPv3HM} that
	include both spin-precession and higher multipole harmonics.
	
	Many previous works have shown higher harmonics to be important for unbiased parameter estimation for high-mass ratio systems~\cite{Varma:2016dnf, bustillo2016impact, Capano:2013raa, Littenberg:2012uj,Bustillo:2016gid, Brown:2012nn, Varma:2014, Graff:2015bba, Harry:2017weg, chatziioannou2019properties, Shaik:2019dym}. 
	A recent study by \citet{kumar2019constraining} found that even for events with mass-ratio close to unity (e.g. GW150914 \cite{abbott2016observation} and GW170104 \cite{scientific2017gw170104}), inclusion of higher harmonic modes in the recovery model provided better constraints for extrinsic parameters such as the binary distance or orbital inclination. 
	Their analysis employs a precessing NR surrogate model \texttt{NRSur7dq2} \cite{blackman2017numerical} which covers mass-ratio $0.5 \le q \le 1.0$\footnote{We define mass ratio as $q\equiv m_2/m_1$, with $m_1 \geq m_2$.}. 
	Recently, the \texttt{NRSur7dq2} model has been extended to cover a larger range of mass ratios $0.16 \le q \le 1.0$~\cite{varma2019surrogate}.
	The upgraded \texttt{NRSur7dq4} model is both fast-to-evaluate and essentially as accurate as the underlying NR waveform data on which the model is built ($0.25 \le q \le 1.0$), while continuing to provide high-accuracy waveforms up to mass ratios of $0.16$.
	It is therefore timely to reanalyze the GW190412 observation using \texttt{NRSur7dq4} to investigate whether using a model that incorporates higher modes and the effects of precession more accurately than other models will change any of the key results reported by the LVC.

	Interestingly, the LVC analysis with \texttt{SEOBNRv4PHM} and \texttt{IMRPhenomPv3HM} waveform models (both having higher harmonics and precession) provide different estimates for various quantities such as the mass ratio and various spin-related parameters.
	For example, while \texttt{IMRPhenomPv3HM} favors mass-ratio $q\sim 1/3$, \texttt{SEOBNRv4PHM} pushes the value further to $q\sim 1/4$ with the estimate for mass-ratio using the aligned spin \texttt{NRHybSur3dq8} model falling in-between. 
	A key aim of our work is to analyze GW190412 data with the \texttt{NRSur7dq4} model to help to reconcile the conflicting parameter estimates.
	We further use two different Phenom models -- \texttt{IMRPhenomPv3HM} and \texttt{IMRPhenomXPHM} \cite{Pratten:2020ceb} -- to support our results.  

	The rest of the paper is organized as follows.
	Section \ref{Sec:Bayes} presents a brief outline of the data analysis framework including Bayesian parameter estimation.  
	In Section \ref{Sec:RealData}, we re-anlyze GW190412 strain data using \texttt{NRSur7dq4} and provide a comparison with results obtained from two phenomenological models, (\texttt{IMRPhenomPv3HM} and \texttt{IMRPhenomXPHM}). 
	We find that an analysis with \texttt{NRSur7dq4} provides a tighter constraint on the mass-ratio than analyses using the \texttt{IMRPhenomPv3HM} and \texttt{IMRPhenomXPHM} models while favoring a broader posterior for the spin precession parameter. 
	We further constrain the binary to be more face-on.
	In this section, we also investigate the effects of higher order harmonics while inferring source properties for GW190412 event. 
	We show that even though $\ell=4$ harmonic modes have negligible signal-to-noise ratio, omission of these modes will influence the inferred posterior distribution of several source properties such as chirp mass, effective inspiral spin, luminosity distance, and inclination. 
	Finally, in Section \ref{Sec:Conclusion}, we summarize our results and discuss the implications of our findings.
	In Appendix \ref{app:InjRec} we report on parameter estimation results with synthetic GW signals using the \texttt{NRSur7dq4} model.
	
	%==========================================================================
	%==========================================================================
	%==========================================================================
	\section{Data Analysis Framework} 
	\label{Sec:Bayes}
	In this section, we provide an executive summary of the technical details used in our analysis of the source properties of GW190412.
	\subsection{Bayesian Inference}
	The measured strain data in a GW detector, 
	\begin{equation}
	d(t)=h(t;\theta)+n(t),
	\end{equation}
	is assumed to be a sum of the true signal, $h(t)$, and random Gaussian and stationary noise, $n(t)$ only.
	Here $\theta$ is the 15-dimensional set of parameters, divided for convenience into intrinsic and extrinsic parameters, that describe the BBH signal that is embedded in the data.
	The intrinsic parameters are the masses of the two BH components $m_1$ and $m_2$ (with $m_1 \ge m_2$), dimensionless BH spin magnitudes $\chi_1$ and $\chi_2$, and four angles describing the spin orientation relative to the orbital angular momentum. 
	Extrinsic parameters include luminosity distance $D_L$, the sky position $\{\alpha,\delta\}$, binary orientation $\{\theta_{JN},\psi\}$, and phase and time at the merger $\{\phi_0,t_0\}$
	\footnote{We further incorporate data-calibration parameters to account for the uncertainties in the measured strain using procedure described in~\cite{SplineCalMarg-T1400682, Romero-Shaw:2020owr} and prior information about the calibration uncertanties avaliable from ~\cite{GW190412_PE_release}.}.
	
	Given the time-series data $d(t)$ and a model for GW signal $H$, we use Bayes' theorem to compute the \textit{posterior probability distribution} (PDF) of the binary parameters,
	\begin{equation}
	p(\theta | d, H) = \frac{\pi(\theta | H) \mathcal{L}(d|\theta, H)}{\mathcal{Z}(d | H)},
	\end{equation}
	where $\pi(\theta | H)$ is the \textit{prior} astrophysical information of the probability distributions of BBH parameters $\theta$ and $\mathcal{L}(d|\theta, H)$ is the likelihood function describing how well each set of $\theta$ matches the assumptions of the data. 
	$\mathcal{Z}(d|H)$ is called the model evidence or marginalized likelihood:
	\begin{equation}
	\mathcal{Z}(d|H) = \int d\theta \pi(\theta | H) \mathcal{L}(d|\theta, H).
	\end{equation}
	The posterior PDF is the target for parameter estimation, while the evidence is the target for hypothesis testing (sometimes referred to as model selection).
	The likelihood is defined as
	\begin{equation}
	\mathcal{L}(d|\theta,H) \propto \mathrm{exp}\left( - \frac{1}{2}\langle d-h_H(\theta)|d-h_H(\theta) \rangle \right),
	\end{equation}
	under the assumption of Gaussian noise $n(t)$.
	Here, $h_H(\theta)$ is the signal waveform generated under the chosen GW model $H$, and $\langle a | b\rangle$ is the noise-weighted inner product defined as
	\begin{equation}\label{eq:inner_product}
	\langle a | b\rangle := 4 \Re\int_{f_{\rm low}}^{f_{\rm high}} \dfrac{a(f) b^*(f)}{S_n(f)} df,
	\end{equation}
	with $S_n(f)$ being the one-sided power spectral density (PSD) of the detector noise and $*$ represents the complex conjugate. 
	The integration limits $f_{\rm low}$ and $f_{\rm high}$ is chosen to reflect the sensitivity bandwidth of the detectors. 
	Depending on the analysis, $f_{\rm low}$ is set to either 20 Hz or 40 Hz as described below, while $f_{\rm high}$ is set to the Nyquist frequency corresponding to a sampling rate of 4096 Hz.

	To compute the posterior probability distribution of BBH parameters $p(\theta | d, H)$, we use the Bayesian inference package \texttt{parallel-bilby} \cite{ashton2019bilby,smith2019expediting,Romero-Shaw:2020owr} with \texttt{dynesty} \cite{speagle2020dynesty} sampler. 
	We report the sampler configuration settings in Table \ref{Tab:sampler_params}\footnote{Our configuration files and posterior data is made publicly available here: \cite{pe_data}.}. 
	We obtain the GW190412 strain and PSD data for all three detectors (LIGO-Hanford, LIGO-Livingston and Virgo) from the Gravitational Wave Open Science Center~\cite{GWOSC_GW190412, Abbott:2019ebz}. 
	The PSDs for these data were computed through the on-source \texttt{BayesWave} method~\cite{Littenberg:2014oda,Cornish:2014kda,Chatziioannou:2019zvs}, and use the inferred median PSDs in our analysis following the same assumptions as in Ref.~\cite{ligo2020gw190412}.
	To generate the \texttt{NRSur7dq4} waveforms, we use \texttt{gwsurrogate} python package~\cite{gwsurrogate-repo, Field:2013cfa}. 
	\texttt{IMRPhenomPv3HM} and \texttt{IMRPhenomXPHM}, on the other hand, have directly been used from \texttt{LALSuite} software library~\cite{lalsuite}.

	% ------------------------------------------------------------------------------------------------------
% ------------------------------------------------------------------------------------------------------
% ------------------------------------------------------------------------------------------------------
\begin{table}
	\centering
	\begin{tabular}{p{0.22\textwidth}|p{0.22\textwidth}}
		\toprule
		Sampler                  & Parameters          \\
		\hline
		\multirow{3}{*}{\texttt{Dynesty}~\cite{speagle2020dynesty}} & live-points $=1200$ \\
                         & tolerance $=0.1$    \\
                         & nact $=50$ \\
		\botrule	
	\end{tabular}
	%}
	\caption{Benchmark sampler configuration parameters. 
	Values were chosen based on a combination of their recommended default parameters~\cite{ashton2019bilby,smith2019expediting} and private communication with the \texttt{Bilby} development team. Full sampler specifications can be found in the data accompanying this paper~\cite{pe_data}. }
	\label{Tab:sampler_params}
\end{table}
% ------------------------------------------------------------------------------------------------------
% ------------------------------------------------------------------------------------------------------
% ------------------------------------------------------------------------------------------------------

	\subsection{Waveform Models}
	In this work, we primarily use the recently developed numerical relativity-based, time-domain surrogate waveform model \texttt{NRSur7dq4}~\cite{varma2019surrogate}. 
	Numerical relativity surrogates use numerical relativity waveforms as training data and build a highly-accurate ``interpolatory'' model over the parameter space using a combination of reduced-order modeling~\cite{Field:2013cfa,Field:2011mf}, parametric fits, and non-linear transformations of the waveform data~\cite{Blackman:2017dfb}.
	The \texttt{NRSur7dq4} model is built from 1528 NR simulations~\cite{varma2019surrogate} and spans a 7-dimensional parameter space of spin-precessing binaries. 

	The model has been rigorously trained for mass ratio $0.25\le q\le1.0$ and spins $0.0 \le \chi_{1,2} \le 0.8$.
	Yet it can also be safely extrapolated to regions of the parameter space with $q\ge 0.16$ and $\chi_{1,2} \le 1.0$~\cite{varma2019surrogate}.
	In the analysis of the high-mass BBH GW190521~\cite{Abbott:2020tfl,Abbott:2020mjq}, the model was used without issue for $\chi_{1,2} \ge 0.8$. 
	Comparisons to NR for $q=0.16$ and spins $0.0 \le \chi_{1,2} \le 0.8$ show that the model's accuracy is at least comparable to \texttt{SEOBNRv4PHM} (cf. Figure 10 of Ref.~\cite{varma2019surrogate}) over the extrapolated region.
	This is important as GW190412's mass ratio posterior is expected to have non-negligible support beyond $q\le0.25$.
	
	Median values of mismatch between NR data and the \texttt{NRSur7dq4} model for a stellar mass binary with a total mass similar to GW190412 (i.e. $\sim 30-40$ $M_{\odot}$) is $\mathcal{M}\sim 7 \times 10^{-4}$. 
	This implies that the waveforms produced by \texttt{NRSur7dq4} model and NR simulations are expected to be indistinguishable as long as the network SNR $\rho$ is less than $\sqrt{\frac{D}{2\mathcal{M}}} \sim 71$ \cite{flanagan1998measuring,lindblom2008model,mcwilliams2010observing, Purrer:2019jcp}, where $D$ denotes the number of model parameters \cite{chatziioannou2017analytic} ($D=7$ for \texttt{NRSur7dq4}). 
	We therefore do not expect the \texttt{NRSur7dq4} model to impose any systematic bias while inferring the binary properties of GW190412 which has a network SNR $\sim 20$~\cite{ligo2020gw190412}.

	One important limitation of the \texttt{NRSur7dq4} model is its frequency coverage. 
	The model is only able to generate relatively short waveforms, about 20 orbits before merger, making it difficult to analyze low mass systems with a lower cut-off frequency $f_{\rm low}=20$ Hz (current default for most LVC analyses). 
	Generally, at any particular time before merger, different spherical harmonic modes are at different frequencies with higher order modes being at higher frequencies than the dominant $l=2, m=\pm2$ mode.
	For GW190412, not even the dominant modes start at 20 Hz within the restricted numbeer of orbits before merger.
	For the typical masses and spins we encounter in the parameter estimation runs, the \texttt{NRSur7dq4} model's dominant mode starts below $40$ Hz, and so we select $f_{\rm low}=40$ Hz in our studies using \texttt{NRSur7dq4}.
	In the next section, we demonstrate that doing so has only marginal effects on our parameter recovery. 
	While some of the higher modes (e.g. the $(4,4)$ mode) can start above $40$ Hz, 
	the impact from the missing lower frequencies is, however, expected to be negligible~\cite{pierro2016directly} for these subdominant modes. 
	As \texttt{NRSur7dq4} is a time-domain waveform model, we further need to perform a Fourier transform to obtain frequency domain waveforms which could be used for Bayesian parameter estimation. 
	To reduce the effect of Gibbs phenomena, we taper all waveforms at their start using a Tukey window \cite{tukey}. 
	Before performing the Fourier transform, if required, waveforms are further zero-padded at the beginning to ensure that the length of the time domain waveform is consistent with the duration of the data used.
	
	We also use two different phenomenological waveform models for precessing BBHs with higher modes to infer the properties of GW190412. 
	The first one is \texttt{IMRPhenomPv3HM}~\cite{khan2019phenomenological,khan2020including}, a model that has been extensively used in several LVC analyses so far~\cite{ligo2020gw190412,Abbott:2020mjq, Abbott:2020khf}. 
	Additionally, we carry out parameter estimation with a recent phenomenological model, \texttt{IMRPhenomXPHM} \cite{Pratten:2020ceb}, which has been found to be in better agreement with numerical relativity simulation data compared to \texttt{IMRPhenomPv3HM}.
	While we do not perform any new parameter estimation runs with the \texttt{SEOBNRv4PHM} model, as these models are computationally expensive to evaluate, we will compare our findings against parameter estimates from the LVC analysis using this model~\cite{ligo2020gw190412, GW190412_PE_release}.
	
	\subsection{Choice of priors}
	\label{SubSec:Priors}
	Our assumptions for the prior PDFs are identical to the LVC analysis of GW190412.
	\begin{itemize}
		\item[i)] We choose uniform priors for the component masses ($5 M_\odot< m_{1} < 60 M_\odot$ and $5 M_\odot < m_{2} < 60 M_\odot$), as defined in the rest frame of the Earth.
		\item[ii)] Uniform priors are also used for the component dimensionless spins ($0.0 \le \chi_1\le 1.0$ and $0.0 \le \chi_2\le 1.0$), with spin-orientations taken as uniform on the unit sphere.
		\item [iii)] The prior on the luminosity distance is taken to be: $P(D_L) \propto D_L^2$, with  $200 \leq D_L \leq 1300$ Mpc.
		\item [iv)] For the orbital inclination angle $\theta_{JN}$, we assume a uniform prior over $0 \leq \cos\theta_{JN} \leq 1$.
		\item [v)] Priors on the sky location parameters $\alpha,\delta$ (right ascension and declination) are assumed to be uniform over the sky with periodic boundary conditions.
	\end{itemize}
	To further ensure that all \texttt{NRSur7dq4} waveforms generated in our parameter estimation analysis starts at or below 40 Hz, for the \texttt{NRSur7dq4} analysis we use a narrower component mass prior of $24 M_\odot< m_{1} < 60 M_\odot$ and $6 M_\odot < m_{2} < 60 M_\odot$. 
	This more restrictive prior will not impact the analysis as the mass posteriors are safely contained within the prior's boundary.
	The choice of spin prior, and its impact on the overall parameter estimation results, was further investigated in Ref.~\cite{Zevin:2020gxf}.
	While the parameter estimates are sensitive to the prior assumptions, the overall results are qualitatively robust with the spin prior used in this study preferred by the data.

%#######################################################################################################
%#######################################################################################################	
\begin{figure*}[htb]
	\centering
	\subfigure[]{\label{fig:2a}
		\includegraphics[scale=0.49]{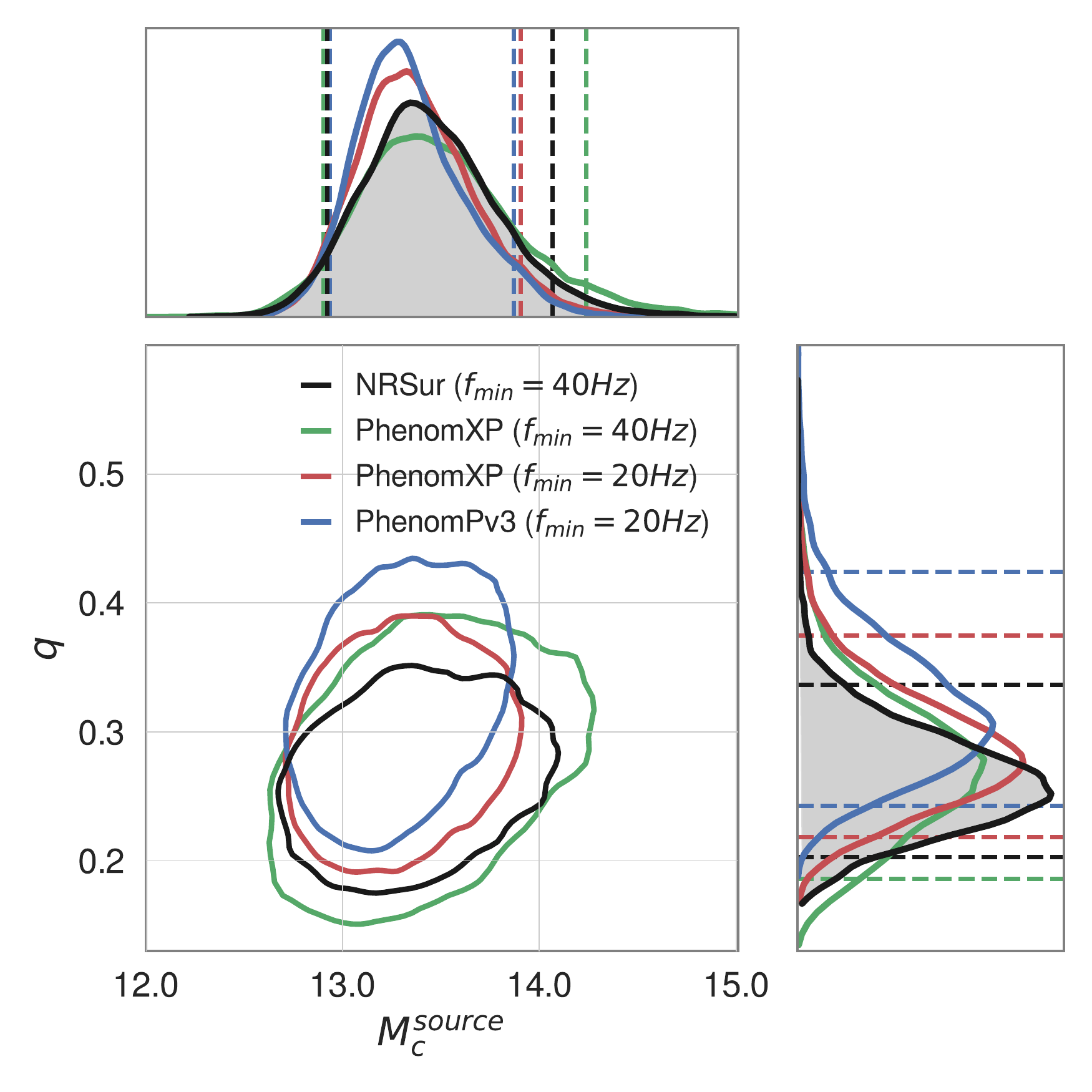}}
	\subfigure[]{\label{fig:2b}
		\includegraphics[scale=0.49]{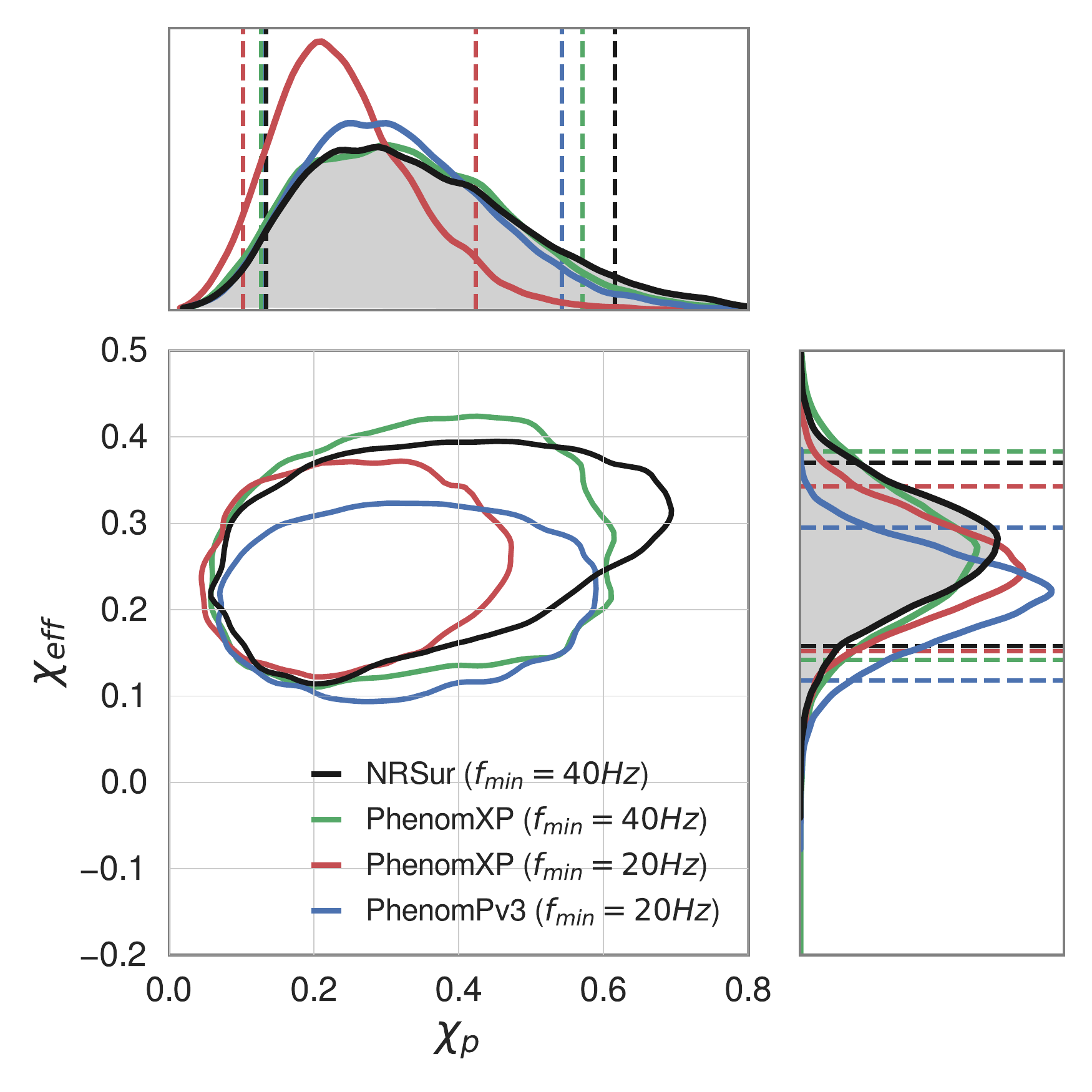}}
	\subfigure[]{\label{fig:2c}
		\includegraphics[scale=0.49]{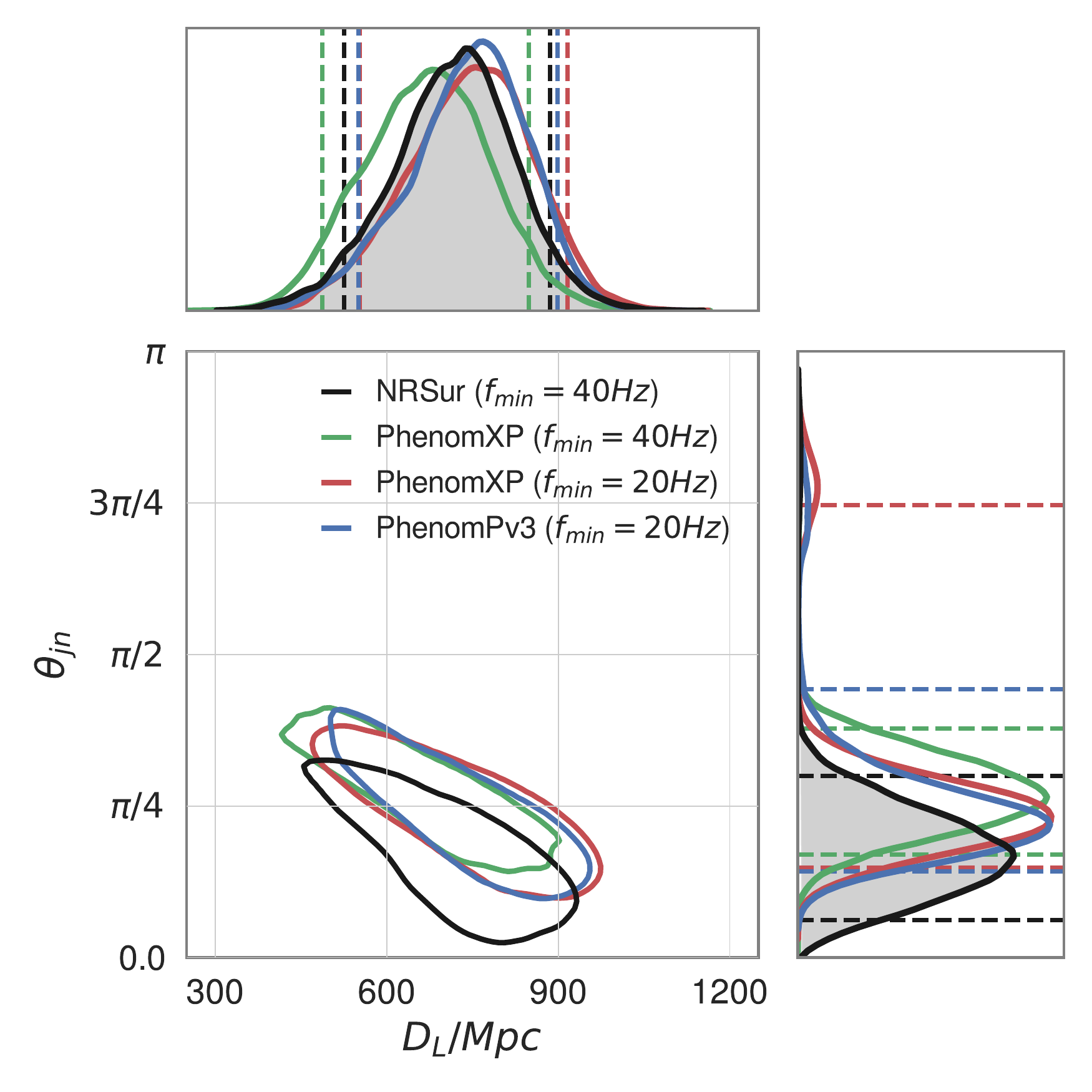}}
	\subfigure[]{\label{fig:2d}
		\includegraphics[scale=0.49]{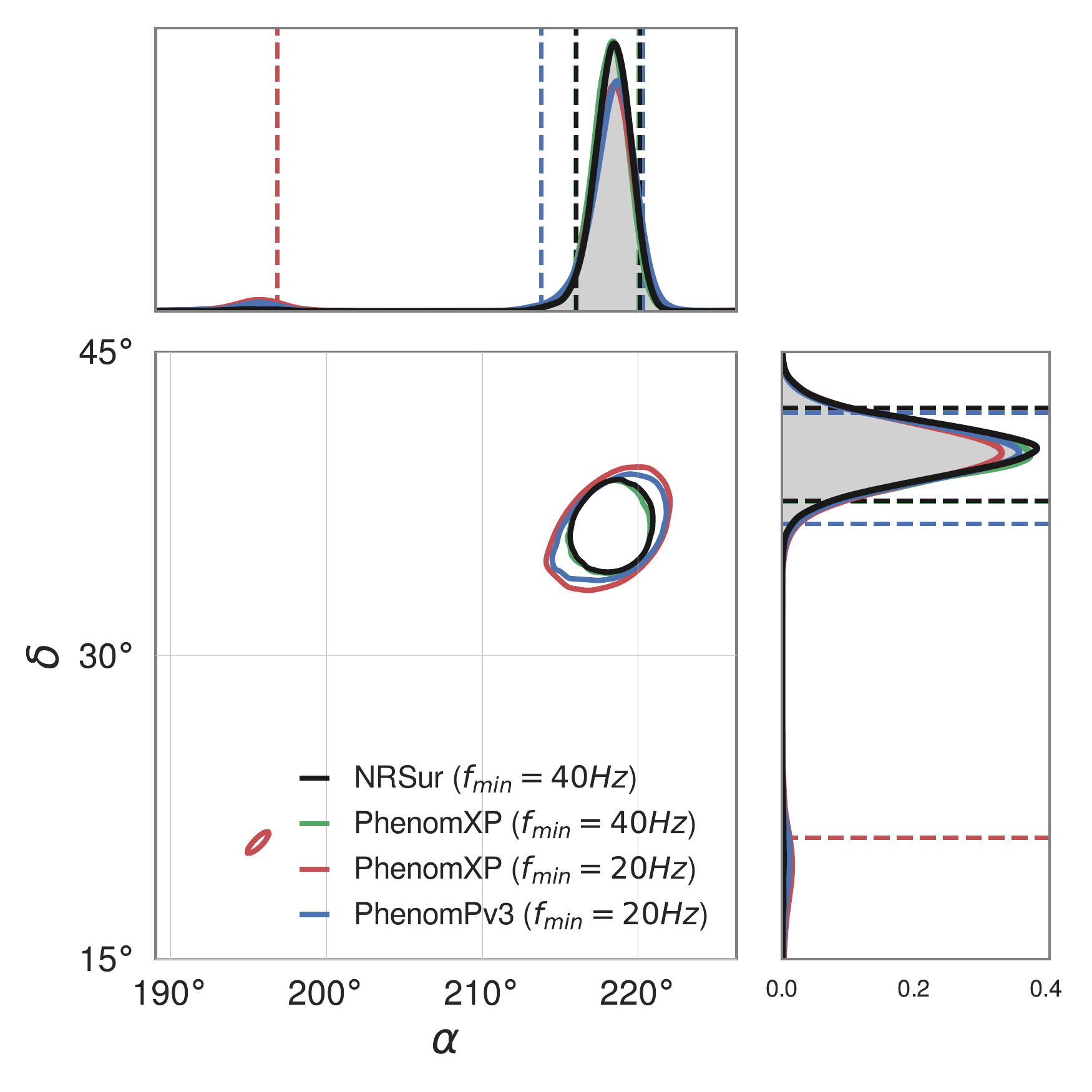}}
	\caption{\label{fig:2} \textbf{Estimated parameters for GW190412 using three different waveform models: \texttt{NRSur7dq4}, \texttt{IMRPhenomXPHM} and \texttt{IMRPhenomPv3HM}}. 
	While the \texttt{NRSur7dq4} analysis is done with lower frequency cutoff $f_{\rm low}=40$ Hz, \texttt{IMRPhenomXPHM} parameter estimation has been carried out with two different values of minimum frequency: $f_{\rm low}=\{20,40\}$ Hz. 
	In panel (a), we show the estimated two-dimensional contours for 90\% confidence interval and one-dimensional kernel density estimates (KDEs) using Gaussian kernel for the source-frame chirp mass $\mathcal{M}^\mathrm{source} (M_\odot$) and mass ratio $q$. 
	Panel (b), (c) and (d) report the corresponding contours and histograms for \{effective inspiral spin $\chieff$, spin precession $\chi_p$\}, \{luminosity distance $D_L$, orbital inclination angle $\theta_{JN}$\} and \{declination $\delta$, right ascension $\alpha$\} respectively. 
	Posteriors for \texttt{NRSur7dq4} and \texttt{IMRPhenomPv3HM} are shown in black and blue receptively. 
	For $\texttt{IMRPhenomXPHM}$, we plot the posteriors in red and green for analysis starting from 20 Hz and 40 Hz respectively. 
	Dashed lines in the one-dimensional panels represent 90\% credible intervals.
    }
\end{figure*}
%#######################################################################################################
%#######################################################################################################

	%==========================================================================
	%==========================================================================
	%==========================================================================
	\section{Analyzing GW190412 Data} 
	\label{Sec:RealData}
	Since the \texttt{NRSur7dq4} model is limited in waveform duration, we are forced to set a higher minimum frequency for the \texttt{NRSur7dq4} analysis of GW190412 in this study as compared to Ref.~\cite{ligo2020gw190412}.
	We choose $f_{\rm low}=40$ Hz and a reference frequency $f_{\rm ref}=60$ Hz whenever analyzing the GW190412 data with \texttt{NRSur7dq4}. 
	Furthermore, for the component masses of the binary, we use a narrower mass prior range as mentioned in Section~\ref{SubSec:Priors}. 
	We note that, for these choices of restricted mass priors, there are no signs of the posterior PDF railing against the prior bounds. 
	The increased value of $f_{\rm low}$ in our analysis, as compared to the LVC analyses~\cite{ligo2020gw190412}, will necessarily incur some loss in the precision of the inferred parameters.
	In addition, we want make direct comparisons between the \texttt{NRSur7dq4} results against analyses with other waveform approximants.
	To ensure a fair and direct approach for these comparisons, we deploy the following sequence of analysis:
    
    \begin{itemize}
    	\item[1.] We first analyze the GW190412 data using \texttt{IMRPhenomPv3HM} with the same priors as used in LVC analysis and $f_{\rm low}=20$ Hz. We set $f_{\rm ref}$ to be 50 Hz.
	These results can then be used as a proxy for the LVC results also using the \texttt{IMRPhenomPv3HM} model~\cite{ligo2020gw190412, GW190412_PE_release}, but generated using our analysis procedure laid out in Section~\ref{Sec:Bayes}. 
	Note that we have used a setup to match the LVC one as best as possible with the \texttt{parallel-bilby} framework.
    	\item[2.] We then repeat the analysis carried out in step 1, but using the new phenomenological waveform approximant \texttt{IMRPhenomXPHM}. 
	Here, we are interested in if using a more modern phenomenological approximant creates any noticeable changes in the inferred parameter PDFs.
    	\item[3.] Next, we perform a parameter estimation using \texttt{IMRPhenomXPHM} waveform model. 
	However, we now change the setup to match the one we intend to use with \texttt{NRSur7dq4}. 
	Namely, (i) the lower cutoff frequency and reference frequency are now set to $f_{\rm low}=40$ Hz and $f_{\rm ref}=60$ Hz respectively, (ii) the mass ratio is constrained to lie within $0.16 \le q \le 1.00$, and (iii) a narrower component mass prior is used as mentioned in Sec.~\ref{SubSec:Priors}.
    	\item[4.] Finally, using the modified setup described in step 3, we redo the analysis with \texttt{NRSur7dq4}.
    \end{itemize}

 	It is important to note that steps 1-3 are designed to investigate whether our results are affected by the choice of a higher value minimum frequency cut-off $f_{\rm low}$ and restricted prior range.
	Results from these four parameter estimations not only allow us to pinpoint any loss in accuracy and precision in inferring parameter properties due to a higher minimum frequency used for the \texttt{NRSur7dq4} analysis, it enables direct comparison to the \texttt{NRSur7dq4} results with the ones obtained using \texttt{IMRPhenomXPHM} within the same analysis setup.
     
	%==========================================================================
	%==========================================================================	
	\subsection{Parameter Inference Results} 
	\label{Sec:PE}
	In Table \ref{tab:parameters}, we present the summary of the inferred source properties for GW190412 employing the \texttt{NRSur7dq4} waveform approximant as well as phenom models (\texttt{IMRPhenomPv3HM} and \texttt{IMRPhenomXPHM}) with different minimum frequencies used in the Bayesian analysis. 
	We report our results as median values and associated one-dimensional 90\% credible intervals.
	We also compute the signal-to-noise Bayes factors~\cite{Veitch:2009hd} 
	\begin{equation}
	\mathcal{B}=\frac{\mathcal{Z_{\rm H}}}{\mathcal{Z_\mathcal{N}}},
	\end{equation}
	where $\mathcal{Z_{\rm H}}$ and $\mathcal{Z_\mathcal{N}}$ denote the evidence for a signal model $H$ and a noise model $N$ respectively.
	The Bayes factor quantifies how much more likely that the data is described by a signal and not by random noise.
	We further report matched-filter signal-to-noise ratio (SNR) recovery (computed using \texttt{pesummary} \cite{Hoy:2020vys,pesummary-repo}) for each of the waveform approximants. 
	Posteriors of select parameters of interest are then shown in Fig.~\ref{fig:2}. 
	Both the \texttt{NRSur7dq4} and \texttt{IMRPhenomXPHM} models yield comparable values for their Bayes factors and SNRs when $f_{\rm low}$ is set to 40 Hz. 
	Similarly, when $f_{\rm low}$ is set to 20 Hz, both \texttt{IMRPhenomPv3HM} and \texttt{IMRPhenomXPHM} recovers the signal with comparable Bayes factors and matched-filter SNR in all detectors. 
	This suggests that any differences in the inferred parameters between models cannot be attributed to differences in the recovered SNR. 

	%==========================================================================
	%==========================================================================	
	\subsubsection{Mass Ratio}	
	We first focus our attention to the mass recovery of different waveform models. 
	We note that the most interesting parameters for GW190412 are its masses.  
	A key result of our paper is the mass-ratio parameter recovery.
	We find that \texttt{NRSur7dq4} favors a BBH signal with $q=0.26_{-0.08}^{+0.06}$, while the \texttt{IMRPhenomPv3HM} model recovers the signal with $0.31^{+0.11}_{-0.07}$. 
	Therefore, the \texttt{NRSur7dq4} estimate for the mass-ratio closely matches the value reported by the LVC obtained with the precessing EOB waveform model \texttt{SEOBNRv4PHM}~\cite{ligo2020gw190412} but is in tension with the LVC's \texttt{IMRPhenomPv3HM} result.
	Interestingly, \texttt{IMRPhenomXPHM} too favors $q\approx 1/4$. 
	Taken together, our result strongly suggests that the progenitor of GW190412 is a $q=1/4$ BBH system and helps resolve the tension between mass-ratio estimates in LVC analysis with \texttt{IMRPhenomPv3HM} and \texttt{SEOBNRv4PHM}. 

	It is interesting to note that the mass-ratio inferred with the \texttt{NRSur7dq4} model is better constrained compared to when obtained using the \texttt{IMRPhenomXPHM} model regardless of whether $f_{\rm low}$ was set to 20 Hz or 40 Hz (Fig.\ref{fig:2a}). 
	In fact, we do not observe any significant difference between the \texttt{IMRPhenomXPHM} mass-ratio posteriors inferred when using $f_{\rm low}=20$ Hz or $f_{\rm low}=40$ Hz. 
	We attribute this improvement in the \texttt{NRSur7dq4} result to not being dependent on the approximations that goes into currently available phenomenological modelling (including \texttt{IMRPhenomXPHM}).
	
	We further note that the higher modes of \texttt{IMRPhenomPv3HM} are not calibrated to NR~\cite{khan2020including}. In addition, typical \texttt{IMRPhenomPv3HM} mismatch for GW190412 like binaries (i.e. $\sim 30-40$ $M_{\odot}$) is $\mathcal{M}\sim  1.5 \times 10^{-2}$ \cite{khan2020including}. This suggests inferred source properties for a GW190412-like binary using \texttt{IMRPhenomPv3HM} are likely to be biased if the newtork SNR is $\ge \sqrt{\frac{D}{2\mathcal{M}}} \sim 16.5$. This could explain the difference in mass-ratio estimates between \texttt{IMRPhenomPv3HM} and \texttt{NRSur7dq4}. 

	%==========================================================================
	%==========================================================================	 
	\subsubsection{Chirp Mass}
	We find that the source-frame chirp mass $\mathcal{M}^\mathrm{source}$ posteriors for both \texttt{IMRPhenomPv3HM} ($13.22^{+0.52}_{-0.35}$) and \texttt{IMRPhenomXPHM} ($13.25^{+0.52}_{-0.29}$) using usual $f_{\rm low}=20$ Hz matches closely with each other (Fig.\ref{fig:2a}). 
	These estimates further match the reported LVC values~\cite{ligo2020gw190412} obtained using \texttt{IMRPhenomPv3HM} and \texttt{SEOBNRv4PHM} models.
	On the other hand, \texttt{NRSur7dq4} yields a broader posterior for $\mathcal{M}^\mathrm{source}$ ($13.27^{+0.60}_{-0.49}$). 
	However, the $\mathcal{M}^\mathrm{source}$ posterior for \texttt{NRSur7dq4} matches very closely with the \texttt{IMRPhenomXPHM} posterior {($13.35^{+0.73}_{-0.51}$) obtained using $f_{\rm low}=40$ Hz, the same value of minimum frequency cut-off set for \texttt{NRSur7dq4} analysis. 
	This indicates that the broadening in posterior for \texttt{NRSur7dq4} can be attributed to the loss of information between 20 Hz and 40 Hz.

	%==========================================================================
	%==========================================================================	
	\subsubsection{Spins}

	GW190412 is the first event with a strong constraint on $\chi_1$, the spin magnitude associated with the more massive BH. 
	In Table \ref{tab:parameters} we report the inferred $\chi_1$ values using \texttt{NRSur7dq4} ($0.48^{+0.26}_{-0.22}$), which mostly matches with the \texttt{IMRPhenomXPHM} analysis with $f_{\rm low}=40$ Hz ($0.47^{+0.21}_{-0.21}$). 
	When setting the value of $f_{\rm low}$ to 20 Hz the \texttt{IMRPhenomPv3HM} model provides a slightly more constrained inference of $\chi_1$. 
	Finally, we note that \texttt{IMRPhenomXPHM} provides a narrower estimation of the dimensionless primary spin magnitude $\chi_1$ ($0.39^{+0.16}_{-0.16}$) than \texttt{IMRPhenomPv3HM} ($0.40^{+0.21}_{-0.22}$) for an analysis that starts from 20Hz. 
	The spin magnitude for the less massive BH remains uninformative for all models considered.

	In Fig.\ref{fig:2b}, we show the recovery of the effective inspiral spin parameter~\cite{Ajith:2009bn, Santamaria:2010yb, Vitale:2016avz},
    \begin{equation}\label{eq:chi_eff}
	\chieff = \frac{m_1 \chi_1 \cos\theta_1 + m_2 \chi_2 \cos\theta_2}{m_1 + m_2} \,,
	\end{equation}
	and the spin precession parameter~\cite{Hannam:2013oca, Schmidt:2014iyl},
	\begin{equation}
	\chi_p = \max\left\{\chi_1 \sin \theta_1, \frac{q(4q+3)}{4+3q} \chi_2 \sin \theta_2\right\} \,,
	\end{equation}
    where $\theta_{1}$ and $\theta_{2}$ are the tilt angles between the spins and the orbital angular momentum respectively.
	We find that $\chieff$ posteriors inferred with \texttt{IMRPhenomPv3HM} model are inconsistent with the ones inferred with \texttt{NRSur7dq4} and \texttt{IMRPhenomXPHM} models. 
	Moreover, the $\chieff$ posteriors obtained using \texttt{NRSur7dq4} and \texttt{IMRPhenomXPHM} models match with each other irrespective of whether $f_{\rm low}$ has been set to 20 Hz or 40 Hz for \texttt{IMRPhenomXPHM}

	We use Jensen-Shannon (JS) divergence~\cite{Lin_ShannonEntropy} to quantify the difference between the one-dimensional marginalized posteriors obtained with different waveform approximants.
	The JS divergence is a general symmetrized extension of the Kullback-Leibler divergence~\cite{kullback1951} with JS divergence values of 0.0 signifying that the posteriors are identical while a JS divergence value of 1.0 would mean the posterior distributions have no statistical overlap at all.
	We find that the JS divergence between $\chi_{\rm eff}$ posteriors obtained from \texttt{IMRPhenomPv3HM} and \texttt{NRSur7dq4} is 0.336. Similarly, the JS divergence between $\chi_{\rm eff}$ posteriors obtained from \texttt{IMRPhenomPv3HM} and \texttt{IMRPhenomXPHM} is 0.304.
	For context, values above $0.15$ are sometimes considered to reflect non-negligible bias~\cite{LIGO-O2-Catalog}, and values near $0.4$ have large, noticeable bias.
	The $\chieff$ posterior PDFs obtained using \texttt{NRSur7dq4} and \texttt{IMRPhenomXPHM} models show better agreement with each other, somewhat irrespective of whether $f_{\rm low}$ has been set to 20 Hz or 40 Hz for \texttt{IMRPhenomXPHM}, producing a JS divergence of 0.13 and 0.07, respectively. 
	The main difference between \texttt{NRSur7dq4} and \texttt{IMRPhenomXPHM} ($f_{\rm low} =40$ Hz) $\chi_{\rm eff}$ posteriors is that the \texttt{NRSur7dq4} one is slightly more constrained than \texttt{IMRPhenomXPHM}.
	This underscores the improvement in waveform modeling techniques in \texttt{IMRPhenomXPHM} over its predecessor \texttt{IMRPhenomPv3HM}.

	As expected, when $f_{\rm low}$ is set to 40 Hz, the $\chi_p$ posterior for \texttt{IMRPhenomXPHM} becomes broader than the one inferred for the case with $f_{\rm low}=20$ Hz. 
	The broadening of the posterior can be attributed to a reduction in the number of resolvable spin-precession cycles~\cite{Hannam:2013oca, Schmidt:2014iyl, Purrer:2015nkh, Fairhurst:2019vut}.
	While this is expected to be the dominant cause of broadening, care is required when comparing $\chi_p$ (and to a lesser extent $\chi_{\rm eff}$) posteriors recovered with different values of $f_{\rm ref}$ since both of these spin quantities are only approximately conserved throughout inspiral.

	Interestingly, the $\chi_p$ posterior obtained using \texttt{NRSur7dq4} provides more support for larger values of $\chi_p$. 
	This is not entirely unexpected, as we do expect discrepancies between \texttt{NRSur7dq4} and \texttt{IMRPhenomXPHM} to increase as the BBH system becomes more asymmetric \cite{Pratten:2020ceb}. 
	We return to this issue in Sec.~\ref{Sec:ModellingInaccuracy}.

	%==========================================================================
	%==========================================================================	
	\subsubsection{Inclination and luminosity distance}
	Another noteworthy aspect of our results in the recovery of luminosity distance and inclination angle  between the line-of-sight and the direction of angular momentum of the BBH system (Fig.\ref{fig:2c}). 	
  	We find that the luminosity distance posterior recovered by \texttt{NRSur7dq4} matches with both 
  	\texttt{IMRPhenomPv3HM} and \texttt{IMRPhenomXPHM} for the cases where $f_{\rm low}=20$ Hz.
  	\texttt{IMRPhenomXPHM} yields smaller distances when $f_{\rm low}$ is set to $40$ Hz.
  	
	\texttt{NRSur7dq4} also favors smaller values of inclination angle suggesting the system is closer to a face-on binary. 
	This smaller value of inclination in \texttt{NRSur7dq4} is potentially the reason for a broader $\chi_p$ posterior as precession is known to be less constrained for face-on binaries~\cite{Chatziioannou:2014coa, Vitale:2014mka, OShaughnessy:2014shr, Mandel:2015spa, Littenberg:2015tpa}.
	Such differences in the inferred luminosity distance and inclination between phenomenological model and NR surrogate models had also been observed while a phenomenological model \texttt{IMRPhenomPv2} (the predecessor of \texttt{IMRPhenomPv3HM}) and an NR surrogate model \texttt{NRSur7dq2} (predecessor of \texttt{NRSur7dq4}) were used for the analysis of GW150914 data~\cite{kumar2019constraining}. 
	In that study, it was shown that the omission of subdominant modes in the \texttt{IMRPhenomPv2} model was responsible for these differences.
	While the \texttt{IMRPhenomXPHM} model includes additional mode content $(\ell,m)=\{(2,\pm1),(3,\pm2),(3,\pm3),(4,\pm4)\}$}, it is still missing many of the modes included in the \texttt{NRSur7dq4} model, i.e. $(\ell,m)=\{(2,\pm0),(3,\pm0),(3,\pm1),(4,\pm0),(4,\pm1),(4,\pm2),(4,\pm3)\}$. 
	We suspect subdominant mode modeling may be responsible for the differences in the inferred luminosity distance and inclination seen here.
	
	%==========================================================================
	%==========================================================================	
	\subsubsection{Source localization}
	In Fig.~\ref{fig:2d}, we show the recovery of the sky location parameters. 
	Both right ascension $\alpha$ and declination $\delta$ inferred using all three different waveform approximants, \texttt{IMRPhenomPv3HM},  \texttt{IMRPhenomXPHM} and \texttt{NRSur7dq4}, with both values of minimum frequencies ($f_{\rm low}=20$ Hz and $f_{\rm low}=40$ Hz) match well with each others. 
	Using \texttt{NRSur7dq4} model in analysis does not provide any extra information for the source's location on the sky, when combined with the distance information shown in Fig.~\ref{fig:2c} the three dimensional volume does however change.
	%==========================================================================
	%==========================================================================	

% ------------------------------------------------------------------------------------------------------
% ------------------------------------------------------------------------------------------------------
% ------------------------------------------------------------------------------------------------------
\begin{table*}
	\caption{Summary of the inferred parameters values for GW190412 (assuming the Planck 2015 cosmology~\cite{Ade:2015xua}). 
	We report the median values with their 90\% credible intervals, obtained using \texttt{NRSur7dq4}, a numerical relativity precessing surrogate model including higher multipoles. 
	For comparison, we also show the results obtained using two different phenomenological models \texttt{IMRPhenomPv3HM} and \texttt{IMRPhenomXPHM}.
	For a direct comparison with the LVC results, we further show the reported parameter values for GW190412 (taken from Table II of~\cite{ligo2020gw190412}).
	Parameter values inferred in this paper using \texttt{IMRPhenomPv3HM} and \texttt{IMRPhenomXPHM} here are broadly consistent with published LVC analyses.
	\footnote{Symbols:  $M/ M_\odot$: Detector-frame total mass; $\mathcal{M}/ M_\odot$: Detector-frame chirp mass; $m_1/ M_\odot$: Detector-frame primary mass; $m_2/ M_\odot$: Detector-frame secondary mass; $M_\mathrm{f}/ M_\odot$: Detector-frame final mass; $M^\mathrm{source}/ M_\odot$: Source-frame total mass; $\mathcal{M}^\mathrm{source}/ M_\odot$: Source-frame chirp mass;  $m_1^\mathrm{source}/ M_\odot$: Source-frame primary mass; $m_2^\mathrm{source}/ M_\odot$: Source-frame secondary mass; $q$: Mass ratio; $\chi_\mathrm{eff}$: Effective inspiral spin parameter; $\chi_\mathrm{p}$: Effective precession spin parameter; $\chi_1$: Dimensionless primary spin magnitude; $\chi_1$: Dimensionless secondary spin magnitude; 
	$D_\mathrm{L}/\mathrm{Mpc}$: Luminosity distance; $\theta_{JN}$: Inclination angle; 
	$\rho_\mathrm{HLV}$: network SNR: network matched-filter SNR for the Hanford, Livingston and Virgo detectors.
	$\ln\mathcal{B}_\mathrm{s/n}$: natural Log Bayes factor;  
    }
	}
% ------------------------------------------------------------------------------------------------------
	\begin{ruledtabular}
		\begin{tabular}{l | c | c c c | c c }			
			&NRSur7dq4 &PhenomPv3HM & PhenomXPHM  &PhenomXPHM &PhenomPv3HM &SEOBNRv4HM\\ 
			& ($f_{\rm low}=40$ Hz) &($f_{\rm low}=20$ Hz) & ($f_{\rm low}=20$ Hz) & ($f_{\rm low}=40$ Hz) & (\texttt{LVC Result}~\cite{ligo2020gw190412}) &(\texttt{LVC Result}~\cite{ligo2020gw190412})\\ 
			\hline
			$M/ M_\odot$  &$45.22^{+4.66}_{-3.80}$ & $42.39^{+3.90}_{-3.44}$     &$43.91^{+4.31}_{-3.51}$   &$44.41^{+6.55}_{-4.08}$  &$42.5^{+4.4}_{-3.7}$ &$45.7^{+3.5}_{-3.3}$\\
			$\mathcal{M}/ M_\odot$ &$15.26^{+0.70}_{-0.49}$ &$15.22^{+0.28}_{-0.20}$ &$15.26^{+0.36}_{-0.21}$ &$15.16^{+0.79}_{-0.39}$ &$15.2^{+0.3}_{-0.2}$ &$15.3^{+0.1}_{-0.2}$\\
		    $m_1/ M_\odot$  &$35.85^{+5.51}_{-4.69}$\ &$32.25^{+4.99}_{-4.84}$ &$34.20^{+5.36}_{-4.78}$ &$34.88^{+7.99}_{-5.41}$ &$32.3^{+5.7}_{-5.2}$ &$36.5^{+4.2}_{-4.2}$\\
		    $m_2/ M_\odot$ &$9.34^{+1.33}_{-1.05}$ &$10.12^{+1.52}_{-1.09}$ &$9.7^{+1.34}_{-1.08}$ &$9.46^{+1.70}_{-1.51}$ &$10.1^{+1.6}_{-1.2}$ &$9.2^{+0.9}_{-0.7}$\\
		    $M_\mathrm{f}/ M_\odot$ &$43.93^{+4.71}_{-3.86}$  &$41.05^{+4.00}_{-3.54}$ &$42.58^{+4.42}_{-3.62}$ &$43.10^{+6.70}_{-4.17}$ &- &-\\
			\rule{0pt}{4ex}%
% ------------------------------------------------------------------------------------------------------
		    $M^\mathrm{source}/ M_\odot$  &$39.50^{+3.84}_{-3.15}$ &$36.86^{+3.21}_{-2.74}$ &$38.15^{+3.67}_{-3.05}$ &$39.10^{+5.48}_{-3.50}$ &$36.9^{+3.7}_{-2.9}$ &$39.7^{+3.0}_{-2.8}$\\
		    $\mathcal{M}^\mathrm{source}/ M_\odot$ &$13.27^{+0.60}_{-0.49}$ &$13.22^{+0.52}_{-0.35}$ &$13.25^{+0.52}_{-0.39}$ &$13.35^{+0.73}_{-0.51}$ &$13.3^{+0.5}_{-0.3}$ &$13.3^{+0.3}_{-0.3}$\\
			$m_1^\mathrm{source}/ M_\odot$ &$31.31^{+4.66}_{-3.97}$ &$28.01^{+4.18}_{-3.98}$ &$29.70^{+4.58}_{-4.07}$ &$30.69^{+6.80}_{-4.62}$ &$28.1^{+4.8}_{-4.3}$ &$31.7^{+3.6}_{-3.5}$\\
			$m_2^\mathrm{source}/ M_\odot$  &$8.15^{+1.19}_{-0.94}$ &$8.78^{+1.47}_{-1.04}$ &$8.42^{+1.25}_{-0.99}$ &$8.32^{+1.58}_{-1.38}$ &$8.8^{+1.6}_{-1.1}$ &$8.0^{+0.9}_{-0.7}$\\
		    $M_\mathrm{f}^\mathrm{source}/ M_\odot$  &$38.36^{+3.92}_{-3.19}$ &$35.66^{+3.29}_{-2.84}$ &$37.00^{+3.74}_{-3.12}$ &$37.94^{+5.60}_{-3.58}$ &$35.7^{+3.8}_{-3.0}$ &$38.6^{+3.1}_{-2.8}$\\
			\rule{0pt}{4ex}%
% ------------------------------------------------------------------------------------------------------
			$q$  &$0.26^{+0.08}_{-0.06}$ &$0.31^{+0.11}_{-0.07}$ &$0.28^{+0.09}_{-0.07}$ &$0.27^{+0.10}_{-0.09}$ &$0.31^{+0.12}_{-0.07}$ &$0.25^{+0.06}_{-0.04}$\\
			\rule{0pt}{4ex}%
% ------------------------------------------------------------------------------------------------------
			$\chi_\mathrm{eff}$ &$0.27^{+0.10}_{-0.11}$ &$0.22^{+0.08}_{-0.10}$ &$0.25^{+0.09}_{-0.10}$ &$0.26^{+0.12}_{-0.12}$ &$0.22^{+0.08}_{-0.11}$ &$0.28^{+0.06}_{-0.08}$\\
			$\chi_\mathrm{p}$   &$0.32^{+0.29}_{-0.19}$  &$0.31^{+0.24}_{-0.17}$ &$0.23^{+0.19}_{-0.13}$ &$0.32^{+0.25}_{-0.19}$ &$0.31^{+0.24}_{-0.17}$ &$0.31^{+0.14}_{-0.15}$\\
			$\chi_1$  &$0.48^{+0.26}_{-0.22}$ &$0.40^{+0.21}_{-0.22}$ &$0.39^{+0.16}_{-0.16}$ &$0.47^{+0.21}_{-0.21}$ &$0.41^{+0.22}_{-0.24}$ &$0.46^{+0.12}_{-0.15}$\\
			$\chi_2$ &$0.51^{+0.43}_{-0.45}$ &$0.47^{+0.46}_{-0.42}$ &$0.47^{+0.46}_{-0.42}$ &$0.48^{+0.46}_{-0.43}$ &$0.63^{+0.32}_{-0.52}$ &$0.48^{+0.41}_{-0.41}$\\
			\rule{0pt}{4ex}%
% ------------------------------------------------------------------------------------------------------
			$D_\mathrm{L}/\mathrm{Mpc}$ &$720.54^{+165.39}_{-194.86}$ &$749.59^{+149.09}_{-199.35}$ &$749.68^{+166.21}_{-196.64}$ &$671.76^{+176.53}_{-184.59}$ &$740^{+150}_{-190}$ &$740^{+120}_{-130}$\\
			$\theta_{JN}$ &$0.54^{+0.40}_{-0.34}$ &$0.72^{+0.67}_{-0.28}$ &$0.75^{+1.59}_{-0.29}$ &$0.85^{+0.34}_{-0.32}$ &$0.76^{+0.40}_{-0.29}$ &$0.71^{+0.23}_{-0.21}$\\
			\rule{0pt}{4ex}%
% ------------------------------------------------------------------------------------------------------
			$\rho_\mathrm{H}$ &$9.15^{+0.21}_{-0.30}$ &$9.35^{+0.19}_{-0.28}$ &$9.23^{+0.17}_{-0.27}$ &$9.11^{+0.19}_{-0.28}$ &$9.5^{+0.2}_{-0.3}$ &$9.5^{+0.1}_{-0.2}$\\
			$\rho_\mathrm{L}$  &$15.18^{+0.20}_{-0.30}$ &$16.02^{+0.21}_{-0.28}$ &$16.03^{+0.20}_{-0.27}$ &$15.22^{+0.22}_{-0.28}$ &$16.1^{+0.2}_{-0.3}$ &$16.2^{+0.1}_{-0.2}$\\
			$\rho_\mathrm{V}$  &$4.02^{+0.26}_{-0.59}$ &$3.53^{+0.35}_{-1.07}$ &$3.58^{+0.27}_{-1.34}$ &$4.03^{+0.25}_{-0.55}$ &$3.6^{+0.3}_{-1.0}$ &$3.7^{+0.2}_{-0.5}$\\
			$\rho_\mathrm{HLV}$ &$18.17^{+0.19}_{-0.30}$ &$18.87^{+0.18}_{-0.28}$ &$18.82^{+0.20}_{-0.31}$ &$18.18^{+0.20}_{-0.30}$ &$19.0^{+0.2}_{-0.3}$ &$19.1^{+0.2}_{-0.2}$\\
			\rule{0pt}{4ex}%
% ------------------------------------------------------------------------------------------------------
			$\ln\mathcal{B}_\mathrm{s/n}$  &$131.99^{+0.24}_{-0.24}$ &$144.49^{+0.21}_{-0.21}$ &$143.38^{+0.24}_{-0.24}$ &$131.92^{+0.23}_{-0.23}$ &- &-\\			
		\end{tabular}
	\end{ruledtabular}
	\label{tab:parameters}
\end{table*}
% ------------------------------------------------------------------------------------------------------
% ------------------------------------------------------------------------------------------------------
% ------------------------------------------------------------------------------------------------------

%#######################################################################################################
%#######################################################################################################	
\begin{figure*}[htb]
	\centering
	\subfigure[]{\label{fig:3a}
		\includegraphics[scale=0.49]{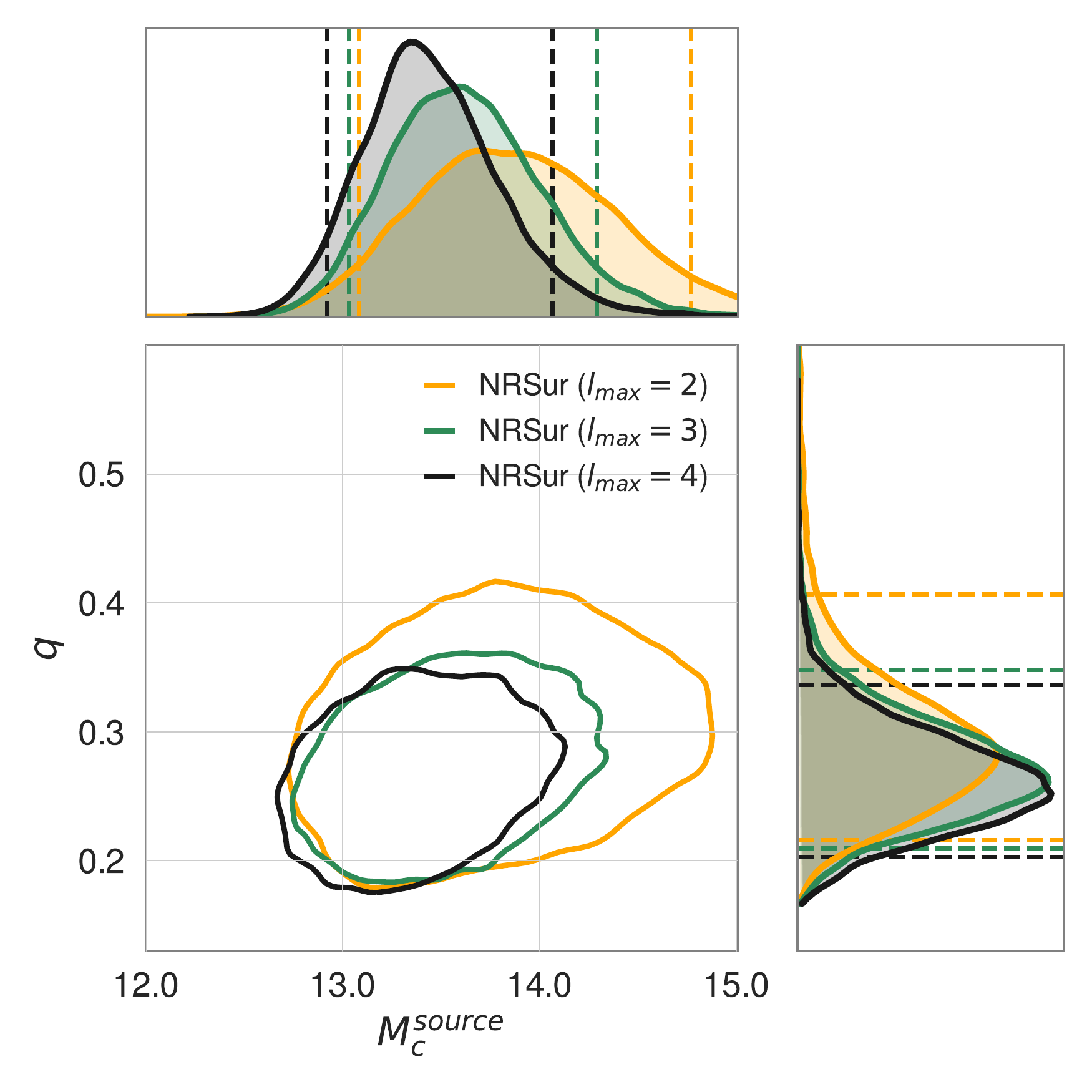}}
	\subfigure[]{\label{fig:3b}
		\includegraphics[scale=0.49]{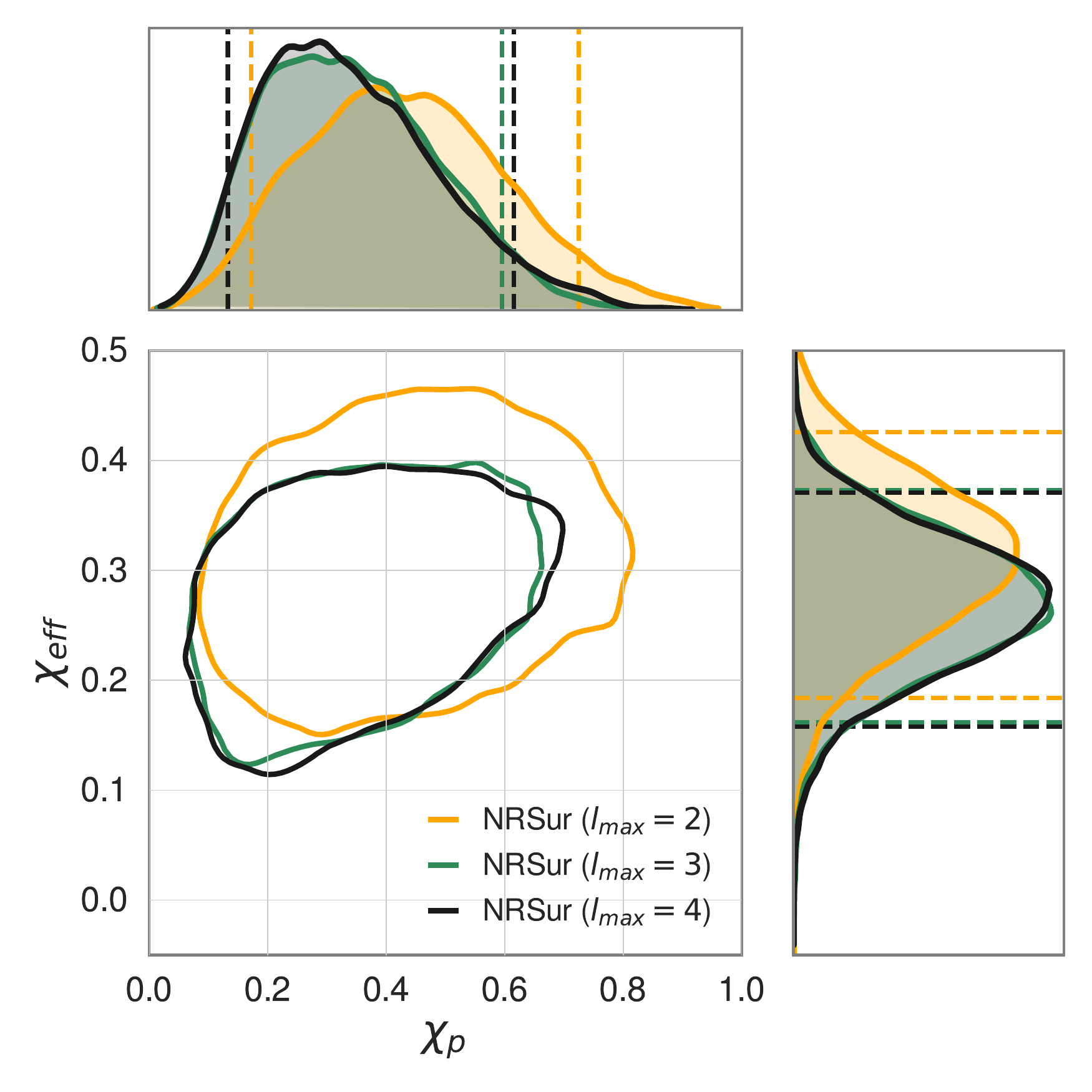}}
	\subfigure[]{\label{fig:3c}
		\includegraphics[scale=0.49]{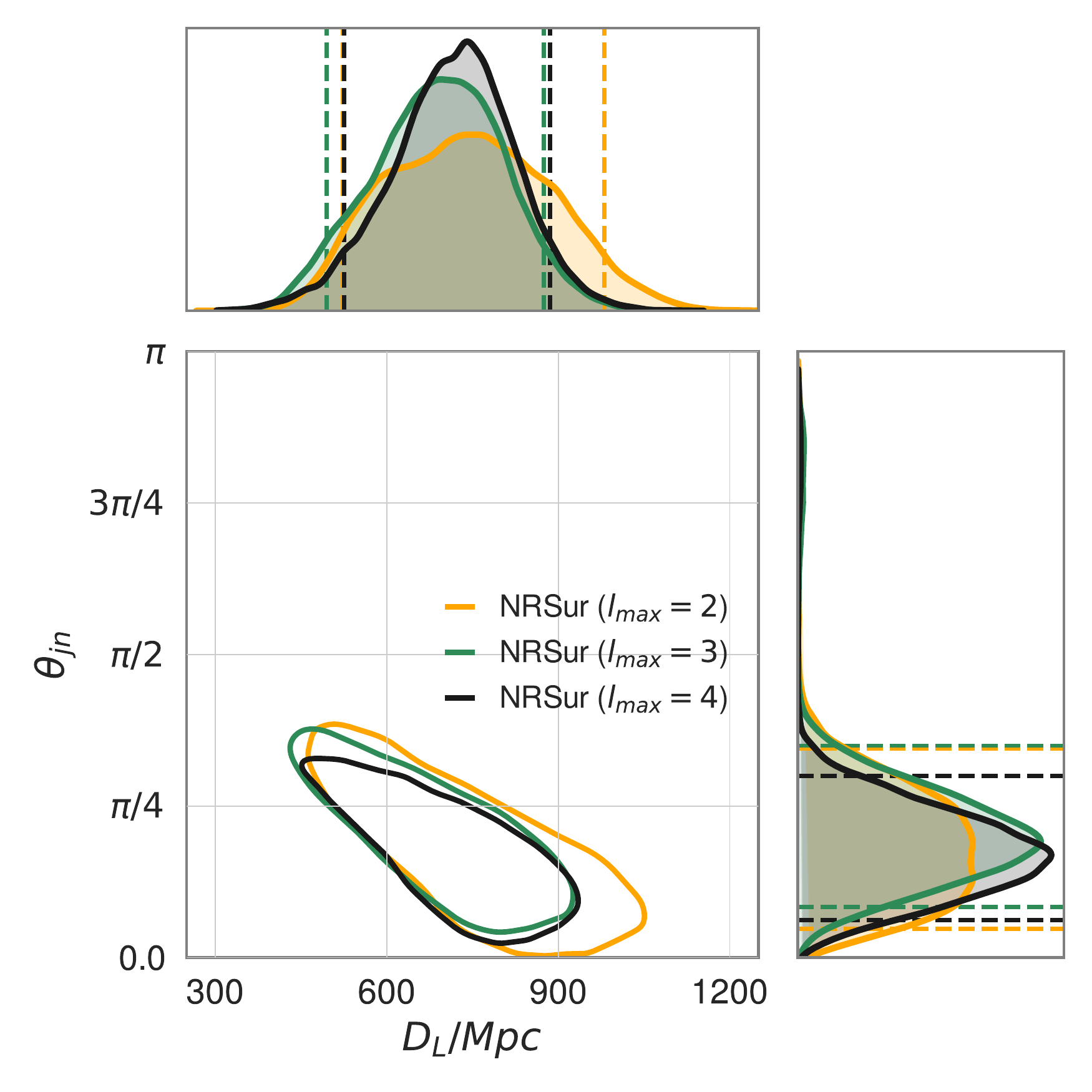}}
	\subfigure[]{\label{fig:3d}
		\includegraphics[scale=0.49]{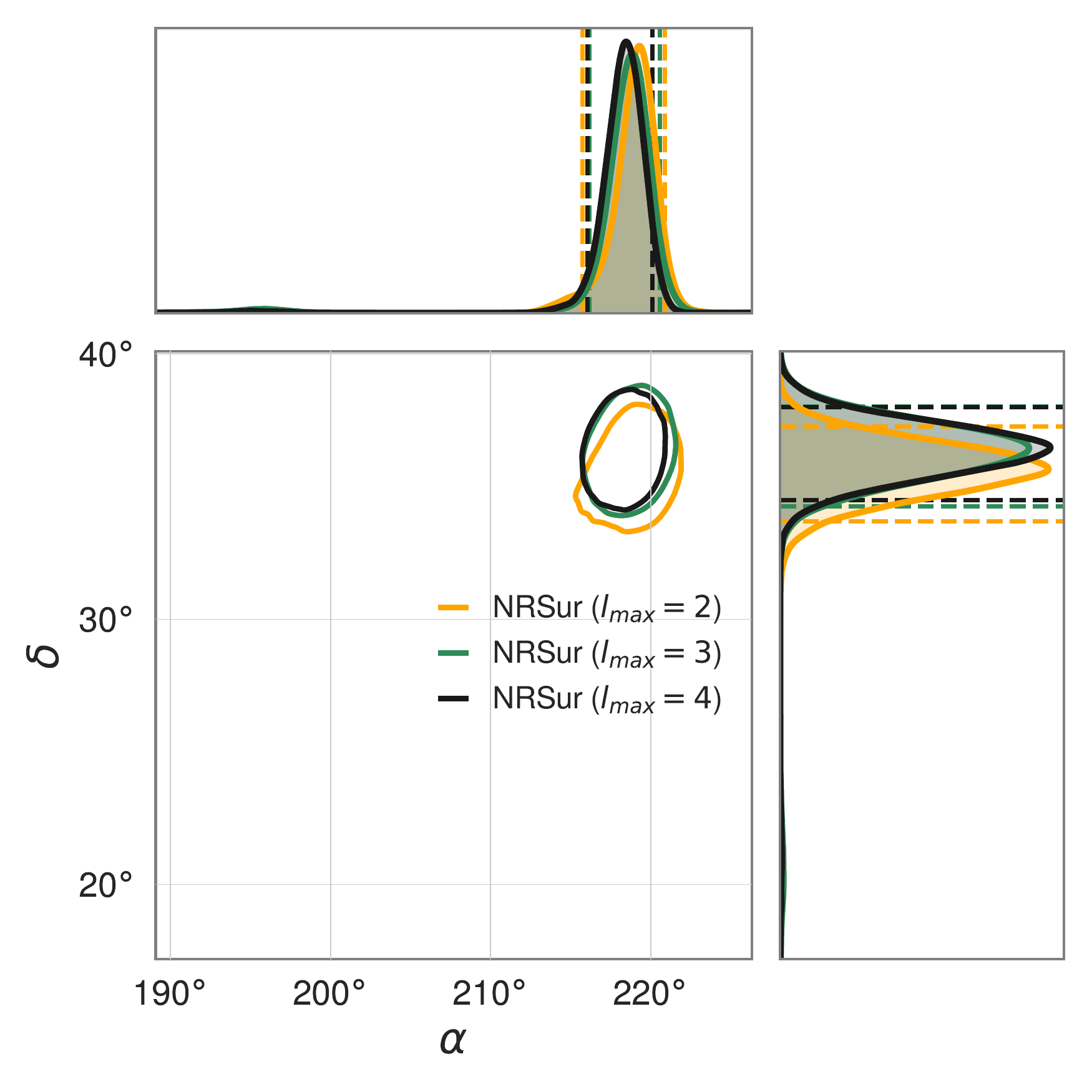}}
	\caption{\label{fig:3} \textbf{Estimated parameters for GW190412 using the \texttt{NRSur7dq4} waveform model varying the spherical harmonic $(l,m)$ mode content} . 
	For all analyses, we use a lower frequency cutoff $f_{\rm low}=40$ Hz. 
	In panel (a), we show the estimated two-dimensional contours for 90\% confidence interval and one-dimensional KDEs using Gaussian kernel for the source-frame chirp mass $\mathcal{M}^\mathrm{source} (M_\odot$) and mass ratio $q$. 
	Panel (b), (c) and (d) report the corresponding contours and histograms for \{effective inspiral spin $\chieff$, spin precession $\chi_p$\}, \{luminosity distance $D_L$, inclination angle $\theta_{JN}$\} and \{declination $\delta$, right ascension $\alpha$\} respectively. 
	Posteriors for \texttt{NRSur7dq4} analysis with all modes $\ell_{max}=2$, $\ell_{max}=3$ and $\ell_{max}=4$ are plotted in orange, green and black lines respectively. 
	Dashed lines in the one-dimensional posterior plots demarcate the 90\% credible regions.}
\end{figure*}
%#######################################################################################################
%#######################################################################################################

% ------------------------------------------------------------------------------------------------------
% ------------------------------------------------------------------------------------------------------
% ------------------------------------------------------------------------------------------------------
\begin{table}
	\caption{Summary of the natural log Bayes factor and network matched filter signal-to-noise ratio (SNR) recovered using NRSur7dq4 model with different all $l\le2$  (i.e. $\ell_{\rm max}=2$), $\ell\le3$ (i.e. $\ell_{\rm max}=3$) and $\ell\le4$ (i.e. $\ell_{\rm max}=4$) modes respectively. 
	The differences in log Bayes factor and SNR recovery demonstrates the importance of higher modes in GW data analysis.
	}
	\footnote{Symbols: $\ln\mathcal{B}_\mathrm{s/n}$: Log Bayes factor;  
		$\rho_\mathrm{HLV}$: network SNR: network matched-filter SNR for the Hanford, Livingston and Virgo detectors.
	}
	\begin{ruledtabular}
		\begin{tabular}{l c c}			
			&$\ln\mathcal{B}_\mathrm{s/n}$ & $\rho_\mathrm{HLV}$\\ 
			\hline
			$\ell_{\rm max}=2$	&$124.65^{+0.22}_{-0.22}$ &$17.54^{+0.26}_{-0.36}$	\\
			$\ell_{\rm max}=3$	&$131.62^{+0.24}_{-0.24}$ &$18.09^{+0.20}_{-0.33}$ \\
			$\ell_{\rm max}=4$	&$131.99^{+0.24}_{-0.24}$ &$18.18^{+0.20}_{-0.30}$
		\end{tabular}
	\end{ruledtabular}
	\label{tab:highermodes}
\end{table}
% ------------------------------------------------------------------------------------------------------
% ------------------------------------------------------------------------------------------------------
% ------------------------------------------------------------------------------------------------------

	%==========================================================================
	%==========================================================================	
	\subsection{Importance of subdominant modes} 
	\label{Sec:HigherModes}
	GW190412 is the first asymmetric mass ratio event detected by LIGO-Virgo collaboration and the first event for which significant SNR support has been observed for a mode other than the dominant quadrupolar mode. 
	The LVC analysis finds considerable SNR in the $\ell=2, m=\pm2$ (SNR $\sim 18.8$) and $\ell=3, m=\pm3$ (SNR $\sim 3.3$) modes~\cite{ligo2020gw190412, Mills:2020thr}. 
	GW190412 therefore provides a prime testing ground to investigate the effects of higher modes in a GW data analysis on an astrophysical BBH observation. 

	In Fig.~\ref{fig:3}, we report the posterior PDFs obtained from analyzing the GW190412 data with the \texttt{NRSur7dq4} model using different modes configurations: $\ell \le2$  (i.e. $\ell_{\rm max}=2$), $\ell \le3$ (i.e. $\ell_{\rm max}=3$) and $\ell\le4$ (i.e. $\ell_{\rm max}=4$) modes respectively including the $m=0$ memory modes.

	In Table~\ref{tab:highermodes}, we report the log Bayes factor and network matched filter signal-to-noise ratio (SNR) recovered using \texttt{NRSur7dq4} model with different mode configurations. 
	We find that even though the difference in SNR recovery between $\ell_{\rm max}=3$ and $\ell_{\rm max}=4$ analysis is small ($\sim1$), including $\ell=4$ modes results in tighter constraints for the chirp mass (Fig.\ref{fig:3a}), the luminosity distance (Fig.\ref{fig:3c}), and the inclination angle (Fig.\ref{fig:3c}).
	In some of the joint posteriors shown in Fig.~\ref{fig:3} we also find that the posterior recovered with the $\ell_{\rm max} = 2$ \texttt{NRSur7dq4} model shows evidence for a secondary peak widely separated from the primary one. 
	Similar spurious peaks were observed in mock parameter estimation studies with heavy BBH systems~\cite{Shaik:2019dym}.
	The inference in the source localization parameters, namely, declination $\delta$ and right ascension $\alpha$ are not affected by the exclusion of $\ell=4$ modes (Fig.\ref{fig:3d}).

% ------------------------------------------------------------------------------------------------------
% ------------------------------------------------------------------------------------------------------
% ------------------------------------------------------------------------------------------------------
\begin{table}
	\caption{Summary of the JS divergence values between the one-dimensional marginalized posterior PDFs obtained using \texttt{NRSur7dq4} with different mode configurations: $\ell_{\rm max}=2$, $\ell_{\rm max}=3$, and $\ell_{\rm max}=4$.
	}
	\footnote{Symbols:  $\mathcal{M}^\mathrm{source}/ M_\odot$ : Source-frame chirp mass; $q$: Mass ratio;  $\chi_\mathrm{eff}$: Effective inspiral spin parameter; $\chi_\mathrm{p}$: Effective precession spin parameter; $D_\mathrm{L}/\mathrm{Mpc}$: Luminosity distance; $\theta_{JN}$: Inclination angle; $\alpha$: right ascension; $\delta$: declination.
	}
	% ------------------------------------------------------------------------------------------------------
	\begin{ruledtabular}
		\begin{tabular}{c c c}
			&JS-Divergnece &JS-Divergence\\
			&between &between\\ 			
			&$\ell_{\rm max}=2$/$\ell_{\rm max}=4$\ &$\ell_{\rm max}=3$/$\ell_{\rm max}=4$\\
			\hline
			$\mathcal{M}^\mathrm{source}/ M_\odot$ &0.34 &0.16 \\
			$q$ &0.21 &0.07\\
			\rule{0pt}{4ex}%
			% ------------------------------------------------------------------------------------------------------
			$\chi_\mathrm{eff}$ &0.22 &0.02 \\
			$\chi_\mathrm{p}$   &0.19 &0.03\\
			\rule{0pt}{4ex}%
			% ------------------------------------------------------------------------------------------------------
			$D_\mathrm{L}/\mathrm{Mpc}$ &0.18 &0.08 \\
			$\theta_{JN}$ &0.13 &0.12\\	
			\rule{0pt}{4ex}%
			% -----------------------------------------------------------------------------------------------------
			$\alpha$  &0.21 &0.11\\
			$\delta$ &0.24 &0.04\\
		\end{tabular}
	\end{ruledtabular}
	\label{tab:HM_JSdivergence}
\end{table}
% ------------------------------------------------------------------------------------------------------
% ------------------------------------------------------------------------------------------------------
% ------------------------------------------------------------------------------------------------------

	To quantify the difference between key posteriors obtained using different mode configurations, we compute the JS divergence~\cite{Lin_ShannonEntropy} between the one-dimensional marginalized posterior.
	In Table \ref{tab:HM_JSdivergence}, we summarize the JS divergence values for between select posterior obtained using different values of $\ell_{\rm max}$. 
	It is evident that not using any higher harmonics beyond $\ell=2$ results in significantly different posteriors for almost all intrinsic and extrinsic parameters. 
	Omitting only $\ell=4$ modes, however, has only marginal effects on the posterior of most parameters, except for the source frame chirp mass and inclination angle. 
	As an $\ell=5$ surrogate model is currently unavailable, we are unable to test whether this family of harmonic mode content would be needed to ensure a sufficiently converged posterior in these two parameters.

	%==========================================================================
	%==========================================================================	
	\subsection{Effect of modelling approximations} 
	\label{Sec:ModellingInaccuracy}
	We now attempt to explain the differences between \texttt{NRSur7dq4} and \texttt{IMRPhenomXPHM} posteriors
	(cf. the $\chieff - \chi_p$ posterior in Fig.~\ref{fig:2b})
	by imposing two \texttt{IMRPhenomXPHM} modeling assumptions (also discussed in \cite{Brugmann:2007zj,Ramos-Buades:2020noq}) into the surrogate. 

\begin{figure}[tb]
	\centering
	\includegraphics[width=\columnwidth]{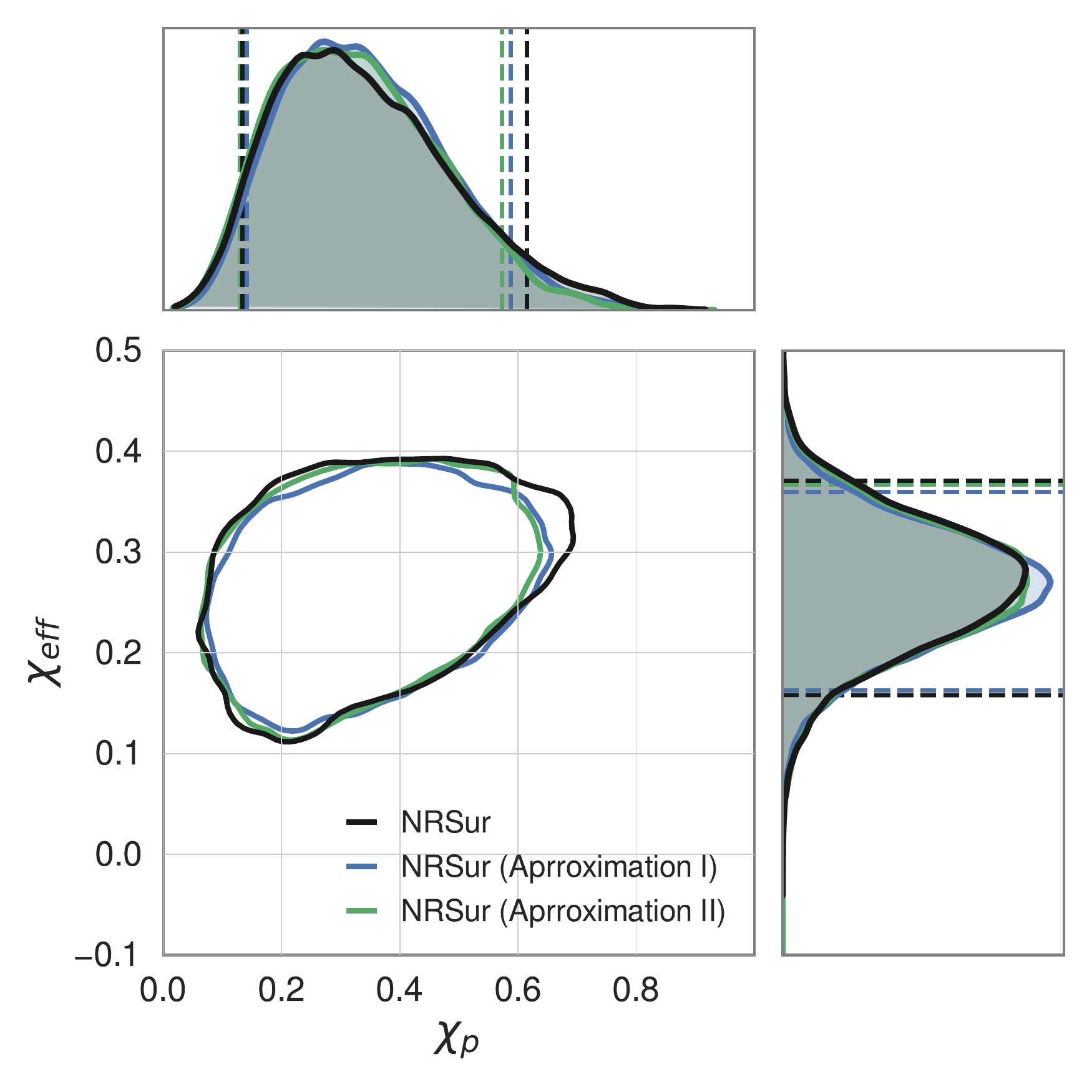}
	\caption{\textbf{Estimated effective inspiral spin$\chieff$ and spin precession $\chi_p$ for GW190412 event using \texttt{NRSur7dq4} waveform.} 
	We show the two-dimensional contours and one-dimensional KDEs using Gaussian kernel of $\chieff$ and $\chi_p$ obtained using the \texttt{NRSur7dq4} waveform model in black. 
	We then show the posteriors using \texttt{NRSur7dq4} after imposing two different modelling approximations that are often assumed for building phenomenological and EOB models. 
	The posteriors recovered (i) after the omission of asymmetric modes is shown in light blue (``Approximation I") and (i) assuming the model reduces to an aligned-spin model in the co-precessing frame is shown in deep blue (``Approximation II"). [Details in text; See Section \ref{Sec:ModellingInaccuracy}]. 
	We find no noticeable difference between posteriors obtained from default \texttt{NRSur7dq4} model and these approximations.}
	\label{fig:mimicphenom}
\end{figure}

	The first modeling assumption made by \texttt{IMRPhenomXPHM} (and many other models) is that the binary black hole system has orbital plane symmetry in the co-precessing frame. 
	This is to say that the asymmetric modes are zero. These modes measure the extent to which the 
	non-precessing formula, $h_{\ell, - m} = (-1)^{\ell} h_{\ell, m}^*$, relating positive and negative $m$ modes is violated in the co-precessing frame. 
	While mode asymmetries of the waveform from precessing systems is a small effect, these features cannot be completely removed with a different choice of frame~\cite{Boyle:2014} and are generally non-zero even in a co-precessing frame.
	In the surrogate model, this approximation can be implemented by setting the asymmetric modes to zero (blue line; Fig.\ref{fig:mimicphenom}). 

	The second modeling assumption made by the phenomological family of models is that in the co-precessing frame the waveform modes are described by an aligned-spin BBH system~\footnote{Instead of using a spin-aligned
	carrier model for the ringdown signal, \texttt{IMRPhenomXPHM} uses quasi-normal modes consistent with
	the remnant mass and spin values from the fully precessing system~\cite{Pratten:2020ceb}. Given GW190412's many in-band orbital cycles, small changes to the ringdown do not play an important roll in checking the twisting approximation used by \texttt{IMRPhenomXPHM}.}.
	In the surrogate model, this approximation can be implemented by setting the in-plane spins in the co-precessing frame to zero (green line; Fig.\ref{fig:mimicphenom}). 
	Note that this second modeling assumption clearly implies the first.

	Neither of these approximations has any appreciable difference on the posteriors, which suggests that other modeling approximations are responsible for the deviations observed in Fig.~\ref{fig:2b}.

%#######################################################################################################
%#######################################################################################################	
%#######################################################################################################
%#######################################################################################################

	%==========================================================================	
	%==========================================================================
	%==========================================================================
	%==========================================================================
	\section{Discussion and Conclusion} 
	\label{Sec:Conclusion}
	In this paper, we used the numerical relativity-based precessing surrogate model, \texttt{NRSur7dq4}, to analyze the BBH signal GW190412 employing a fully Bayesian framework. 
	Restrictive waveform durations in the \texttt{NRSur7dq4} model has forced us to set the lower frequency cutoff to be 40 Hz instead of the usual 20 Hz. 
	Despite this limitation, we demonstrate that an analysis using the \texttt{NRSur7dq4} model is able to efficiently infer binary properties from the observed GW signal. 

	Our analysis broadly agrees with the published LVC results using \texttt{SEOBNRv4PHM} and is in disagreement with the  LVC results using \texttt{IMRPhenomPv3HM}. 
	As such, we believe our results can serve to help resolve the tension between mass-ratio and spin estimates in the official LVC analysis.

	We also find that \texttt{NRSur7dq4} provides improved constraints for the mass-ratio, spin precession, luminosity distance and inclination. 
	Using the \texttt{NRSur7dq4} model we have been able to provide a better constraint on the mass ratio than the state-of-art phenomenological waveform model \texttt{IMRPhenomXPHM}. 
	Furthermore, we show the binary to be more close to a face-on system which results in less constrained estimation of the spin precession parameter. 
	All these result, taken together, indicate that numerical-relativity based surrogate models could help to extract more information of BBH mergers from events like GW190412.
	This is because surrogates have been extensively trained on numerical simulations of precessing systems and include all of the most important subdominant modes.
	We also recommend that future numerical-relativity surrogates should include $\ell_{\rm max} > 4$ harmonic modes as these could be important for the inference of certain parameters (cf.~Sec.~\ref{Sec:HigherModes}).
	
	Another direction of the paper has been to investigate the effects of subdominant modes in parameter estimation with real data instead of synthetic GW datasets.
	Using the \texttt{NRSur7dq4} model we perform parameter inference using a sequence of $\ell_{\rm max} =2,3,$ and $4$ harmonics modes, where $\ell_{\rm max} =4$ corresponds to all modes available in \texttt{NRSur7dq4}.
	We find that, even though the increase in recovered SNR using $\ell=4$ modes is negligible, the omission of subdominant modes can affect the posteriors quite significantly. 
	When using only $\ell_{\rm max} =2$ modes we find that all parameters are biased (cf.~Table~\ref{tab:HM_JSdivergence}) and some posteriors develop spurious secondary peaks. 
	When including $\ell_{\rm max} =3$ modes the chirp mass and inclination angles still show moderate bias. 
	While we suspect $\ell_{\rm max}=4$ modes should be sufficient to resolve the posterior, currently no model has a complete family of $\ell_{\rm max}=5$ modes to check this.
	As more binaries with asymmetric masses are detected, the modelling of higher order modes will become increasingly important even if the SNR contribution from each individual mode is small.
	
	This study also presents a new opportunity to explore how modelling can affect posteriors -- sometimes non-trivially. 
	We note that even though both \texttt{NRSur7dq4} and \texttt{IMRPhenomXPHM} model have higher modes and spin precession, results obtained using these two models do not always agree. 
	Specially, the estimates of mass ratio, spin precession, luminosity distance, and inclination differs depending on which of these two models have been used to analyze the data. 
	The difference in modelling higher order modes and spin effects could potentially be the reason for the observed difference in parameter estimation. 
	We attempted to explain our observed differences by considering two approximations widely used in many effective one body and phenomenological waveform families to model the spin effects: (i) a binary system has orbital plane symmetry in co-precessing frame; and (ii) the gravitational-wave modes
	in the binary's co-precessing frame is described by an aligned-spin system. 
	We find that, for low SNR event like GW190412, such assumptions do not change the posteriors in any significant way. 
	However, with events having higher SNRs, these approximations might yield differences in the recovered posteriors.
	%==========================================================================
	%==========================================================================
	%==========================================================================
	
	\begin{acknowledgments}
		We thank Marta Colleoni, Sascha Husa, Gaurav Khanna, Vijay Varma and Michael Zevin for helpful discussions,
		and Gaurav Khanna and Feroz Shaik for providing technical assistance using the CARNiE cluster.
		CJH acknowledge support of the National Science Foundation, and the LIGO Laboratory.
		LIGO was constructed by the California Institute of Technology and Massachusetts Institute of Technology with funding from the National Science Foundation and operates under cooperative agreement PHY-1764464.
		SEF is partially supported by NSF grants PHY-1806665 and DMS-1912716.
		TI is supported by NSF grants PHY-1806665 and DMS-1912716, and a Doctoral Fellowship provided by UMassD Graduate Studies.
		The computational work of this project was performed on the CARNiE cluster
		at UMassD, which is supported by the ONR/DURIP Grant No.\ N00014181255. 
		A portion of this material is based upon work supported by the National Science Foundation
		under Grant No. DMS-1439786 while the author was in residence at the Institute
		for Computational and Experimental Research in Mathematics in Providence, RI, during the Advances in Computational Relativity program.
		This research has made use of data, software and/or web tools obtained from the Gravitational Wave Open Science Center (https://www.gw-openscience.org), a service of LIGO Laboratory, the LIGO Scientific Collaboration and the Virgo Collaboration. 
		LIGO is funded by the U.S. National Science Foundation. 
		Virgo is funded by the French Centre National de Recherche Scientifique (CNRS), the Italian Istituto Nazionale della Fisica Nucleare (INFN) and the Dutch Nikhef, with contributions by Polish and Hungarian institutes.
		This is LIGO Document Number DCC-P2000384.
	\end{acknowledgments}  
	%==========================================================================

			%==========================================================================
	%==========================================================================
	\appendix
	\section{Parameter Estimation with Synthetic GW Signals} 
	\label{app:InjRec}
	Since the NRSur7dq4 model has not been extensively used in many parameter estimation studies so far, except for analyzing the GW150914 event \cite{varma2020extracting} and the recent GW190521 high-mass BBH~\cite{Abbott:2020mjq}, we perform an injection study to understand potential systematics in greater detail.
	We generate synthetic GW signals with \texttt{NRSur7dq4} model using all available $l \le 4$ modes and inject them in the three-detector LVC network consisting of the LIGO-Hanford, LIGO-Livingston, and Virgo detectors. 
	We then estimate the signal properties using the Bayesian framework described in Section~\ref{Sec:Bayes}. 
	We assume design sensitivity for each of the LIGO-Virgo detectors and use a zero noise configuration.
	
	We choose the injected BBH parameters such that they match the GW190412 properties as inferred from the LVC analysis~\cite{ligo2020gw190412}. 
	We note that the inferred values for the mass-ratio in LVC analysis with \texttt{IMRPhenomPv3HM} and \texttt{SEOBNRv4HM} do not match with each other. 
	While PE with \texttt{IMRPhenomPv3HM} model indicates a mass ratio of $q=0.33_{-0.09}^{+0.12}$, \texttt{SEOBNRv4HM} prefers a more asymmetric mass ratio $q=0.25_{-0.04}^{+0.06}$. 
	We therefore simulate two different synthetic GW signals -- with mass-ratio $q_{\rm inj}=0.25$ and $q_{\rm inj}=0.33$, respectively. 
	This ensures that our injection study is relevant for asymmetric mass-ratio events like GW190412. 
	All other parameter values for both the injections are same. 
	Both the signals were created at a luminosity distance of $D_L=730$ Mpc and with inclination angle $\theta_{JN}=0.73$. 
	We choose dimensionless spin parameters $\chi_1=0.43$ and $\chi_2=0.55$; and spin angles as: $\theta_{1}=1.05$ and $\theta_{2}=1.01$, $\phi_{12}=3.53$ and $\phi_{jl}=3.75$ respectively (cf. Appendix of \cite{Romero-Shaw:2020owr} for definitions of these parameters\footnote{We use $\theta_{1}$ and $\theta_{2}$ to denote the tilt angles}). 
	This corresponds to an effective inspiral spin of $\chieff=0.3$ and spin precession of $\chi_p=0.4$. 
	The chosen sky localization for the injected signals is: right ascension $\alpha=218.29\degree$ and declination $\delta=36.09\degree$. We set $f_{\rm low}=40$ Hz and $f_{\rm ref}=60$ Hz.
		
\begin{figure*}[htb]
	\centering
	\subfigure[]{\label{fig:1a}
		\includegraphics[scale=0.49]{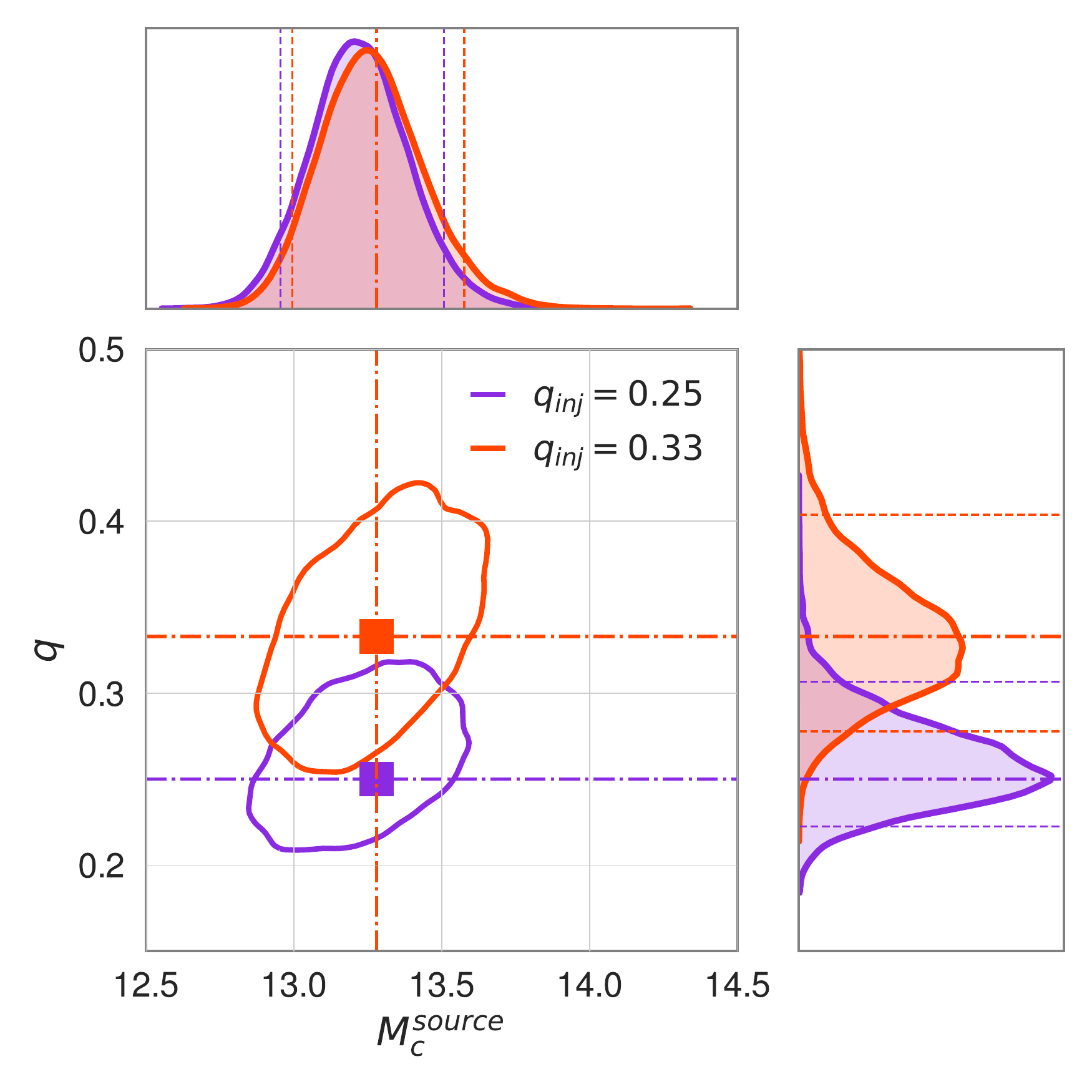}}
	\subfigure[]{\label{fig:1b}
		\includegraphics[scale=0.49]{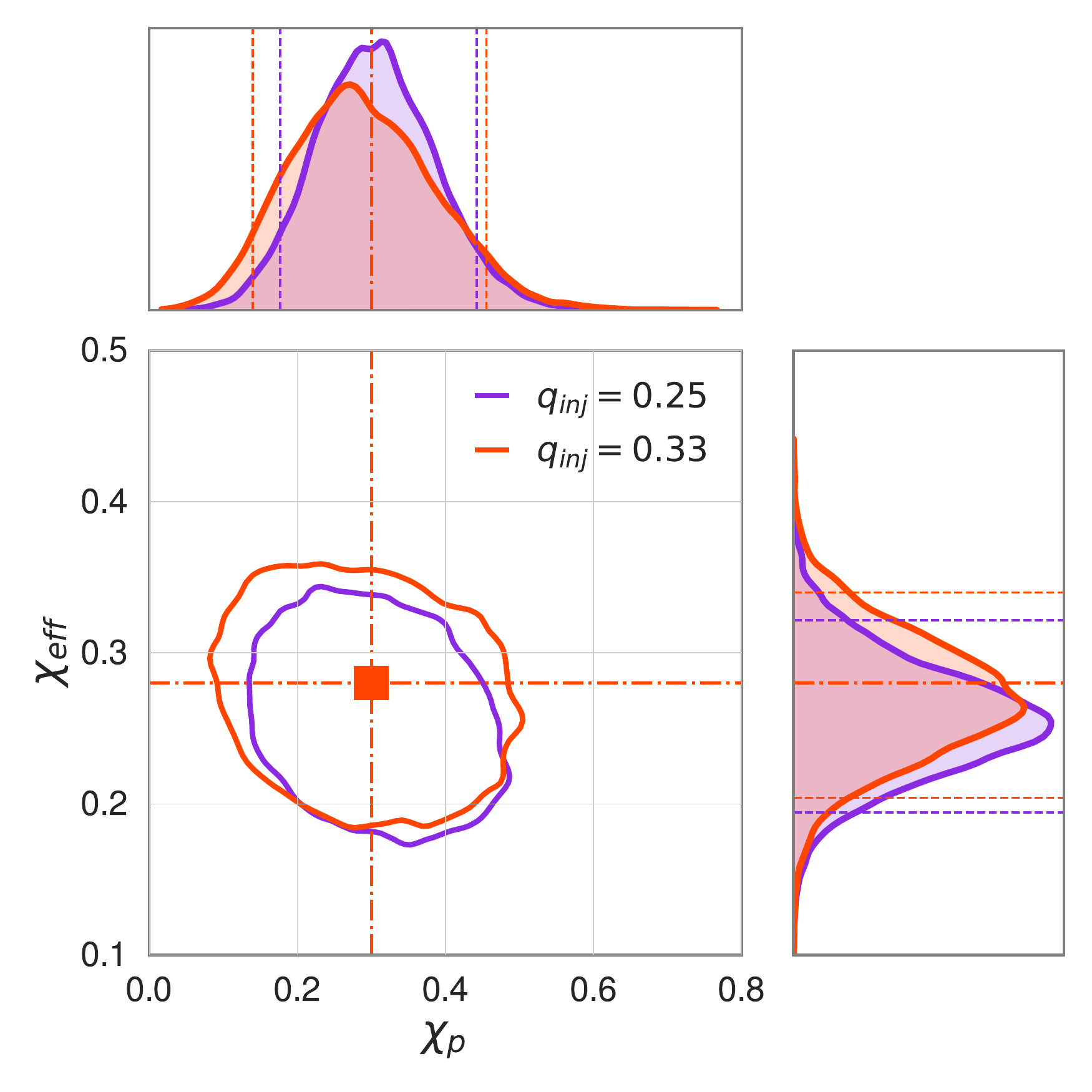}}
	\subfigure[]{\label{fig:1c}
		\includegraphics[scale=0.49]{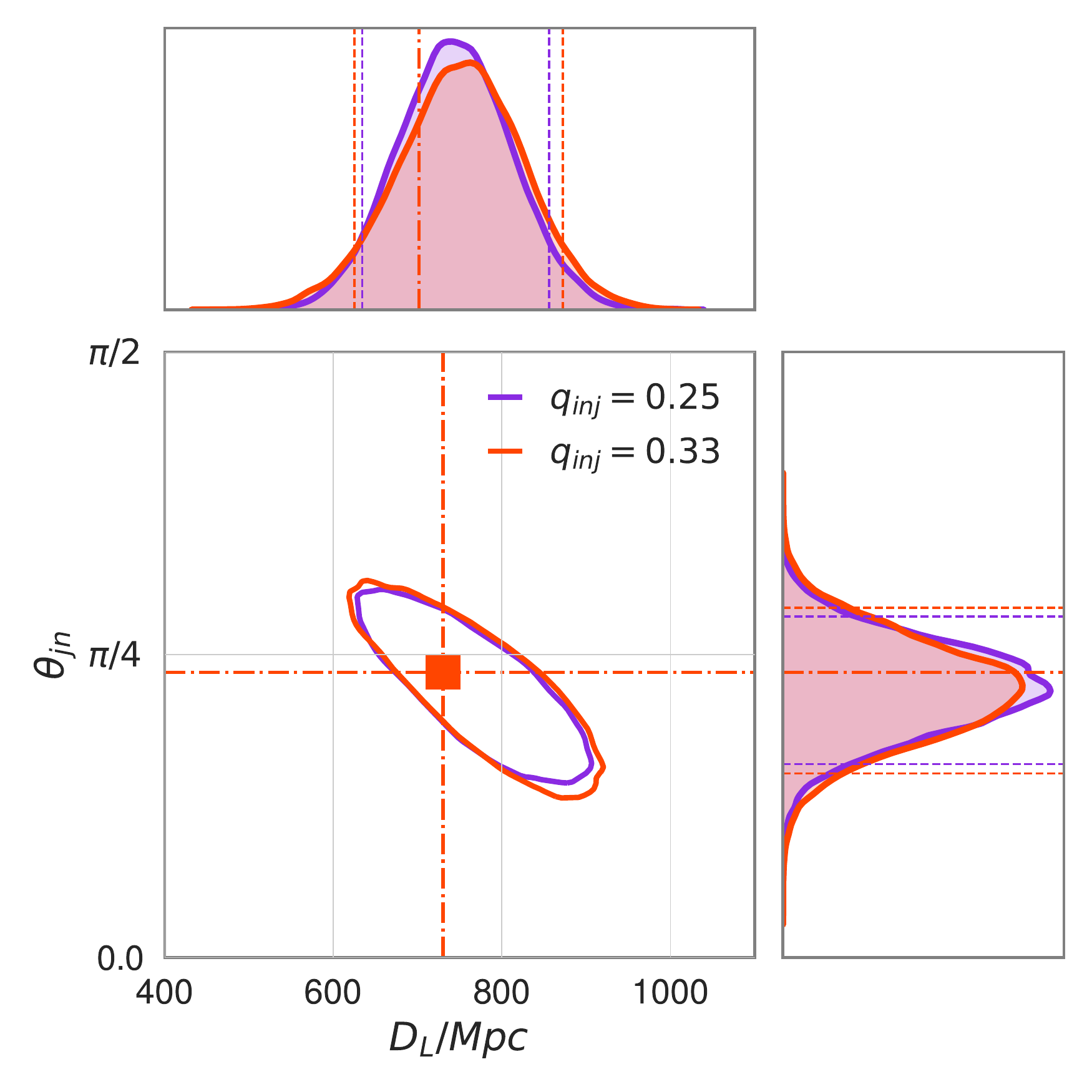}}
	\subfigure[]{\label{fig:1d}
		\includegraphics[scale=0.49]{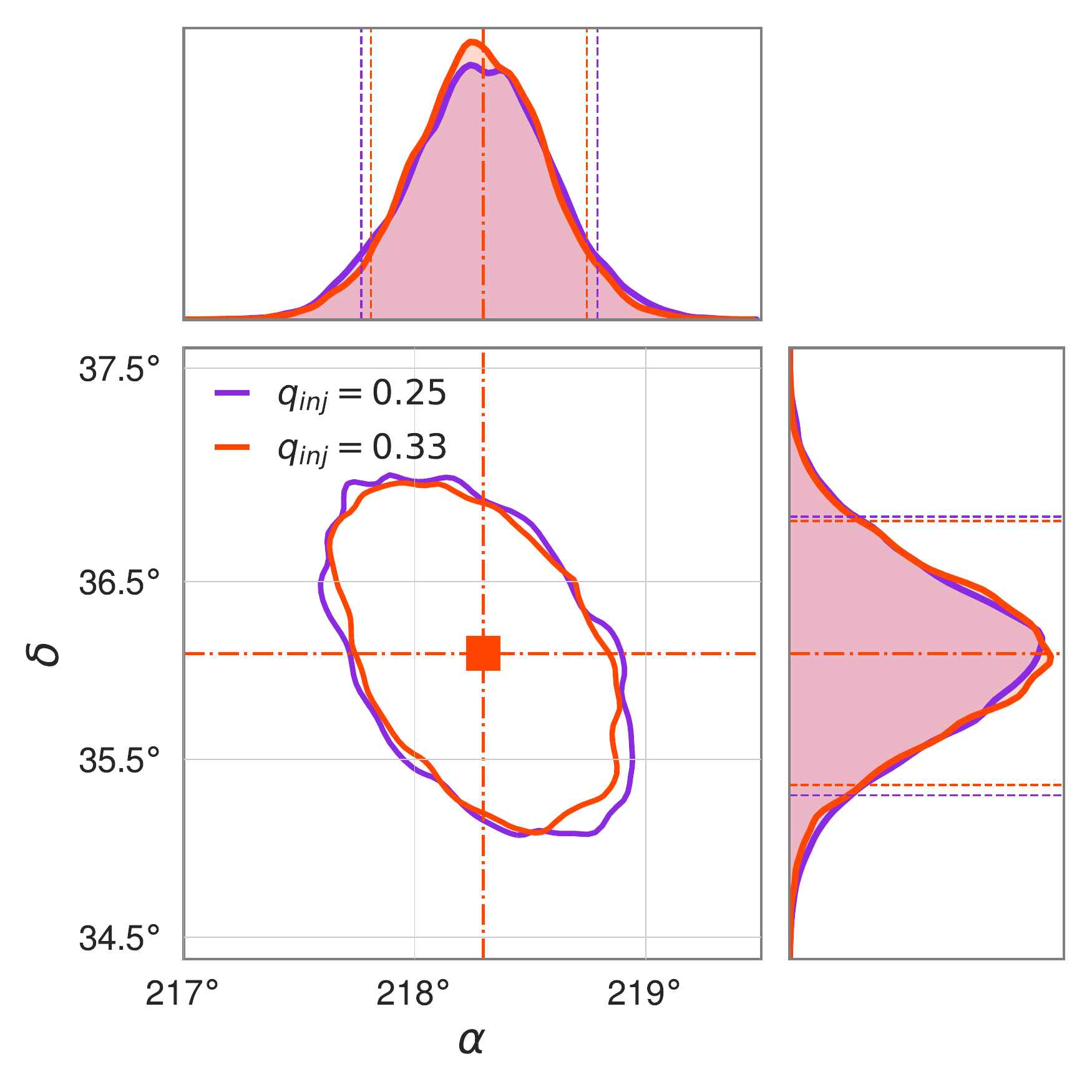}}
	\caption{\label{fig:1} \textbf{Parameter recovery for the synthetic injections with the \texttt{NRSur7dq4} waveform model for two different values of mass ratio: $q_{inj}=0.33$ and $q_{inj}=0.25$.} 
	In (a), we show the estimated two-dimensional contours for 90\% confidence interval and one-dimensional KDEs for the source-frame chirp mass $\mathcal{M}^\mathrm{source} (M_\odot$) and mass ratio $q$. 
	Panel (b), (c) and (d) report the corresponding contours and KDEs using Gaussian kernel for \{effective inspiral spin $\chieff$, spin precession $\chi_p$\}, \{luminosity distance $D_L$, inclination angle $\theta_{JN}$\} and \{declination $\delta$, right ascension $\alpha$\} respectively. 
	Posteriors for $q_{inj}=0.25$ are shown in violet while $q_{inj}=0.33$ posteriors are plotted in orange. 
	In both two-dimensional and one-dimensional panels, dashed dotted lines indicate the true injection values. 
	Dashed lines in the one-dimensional panels represent 90\% credible regions.}
\end{figure*}

	We find that \texttt{NRSur7dq4} model successfully recovers the injected source properties. 
	In Fig.\ref{fig:1}, we show the recovered posteriors for both the injections. 
	Posteriors for $q_{\rm inj}=0.25$ is shown in violet whereas $q_{\rm inj}=0.33$ posteriors are plotted in orange. 
	We find that while both chirp mass and effective spin of the BBH are estimated with comparable precision for both the injections, the mass-ratio and spin precession is well constrained for more asymmetric signal (i.e for $q_{\rm inj}=0.25$). 
	For the extrinsic parameters, we find no significant difference in the recovered posteriors. 
	This injection study demonstrates the efficacy of \texttt{NRSur7dq4} waveform model to successfully recover a true signal from strain data.
	
%#######################################################################################################
%#######################################################################################################	
%#######################################################################################################
%#######################################################################################################
	%==========================================================================
	%==========================================================================
	%%%%%%%%%%%%%%%%%%%%%%%%%%%%%%%%%%%%%%%%%%%%%%%%%%%%%%%%%%%%%%%%%%%%%%%%%%%%%%%
	\section*{References}
	%%%%%%%%%%%%%%%%%%%%%%%%%%%%%%%%%%%%%%%%%%%%%%%%%%%%%%%%%%%%%%%%%%%%%%%%%%%%%%%
	\bibliography{References}

\end{document}